\documentclass[sigconf,natbib=false]{acmart}
\AtBeginDocument{%
  \providecommand\BibTeX{{%
    Bib\TeX}}}

\setcopyright{acmlicensed}
\copyrightyear{2018}
\acmYear{2018}
\acmDOI{XXXXXXX.XXXXXXX}
\acmConference[Conference acronym 'XX]{Make sure to enter the correct
  conference title from your rights confirmation email}{June 03--05,
  2018}{Woodstock, NY}
\acmISBN{978-1-4503-XXXX-X/2018/06}

\usepackage{amsmath,amsfonts}
\usepackage{algorithmic}
\usepackage{graphicx}
\usepackage{textcomp}
\usepackage{xcolor}

\def\BibTeX{{\rm B\kern-.05em{\sc i\kern-.025em b}\kern-.08em
    T\kern-.1667em\lower.7ex\hbox{E}\kern-.125emX}}

\usepackage[
  datamodel=acmdatamodel,
  style=acmnumeric,
  doi=false,
  isbn=false
  ]{biblatex}

\usepackage{multirow}
\usepackage[draft]{zebra-goodies}
\usepackage{float}
\usepackage{flafter}
\usepackage{placeins}

\usepackage{acro}

\usepackage{xspace}

\usepackage[linesnumbered,ruled,vlined]{algorithm2e}

\usepackage[olditem,oldenum]{paralist}
\usepackage{enumitem}
\setlist[itemize]{noitemsep}

\usepackage[most]{tcolorbox}

\usepackage{booktabs}

\usepackage{colortbl}

\usepackage{siunitx}

\usepackage{supertabular}

\usepackage{tabularx}

\usepackage{threeparttable}

\usepackage{pifont} %
\usepackage{fontawesome5}
\usepackage{wasysym} %
\usepackage{amsmath} %
\usepackage{graphicx}
\usepackage{hyperref}

\hypersetup{
    colorlinks=true,
    linkcolor=blue,
    filecolor=magenta,      
    urlcolor=cyan!50!black,
    }

\usepackage{xcolor}

\usepackage[all]{nowidow}

\usepackage{listings}
\usepackage[frozencache,cachedir=_minted-main]{minted}
\makeatletter
\let\old@lstKV@SwitchCases\lstKV@SwitchCases
\def\lstKV@SwitchCases#1#2#3{}
\makeatother
\usepackage{lstlinebgrd}
\makeatletter
\let\lstKV@SwitchCases\old@lstKV@SwitchCases

\lst@Key{numbers}{none}{%
    \def\lst@PlaceNumber{\lst@linebgrd}%
    \lstKV@SwitchCases{#1}%
    {none:\\%
     left:\def\lst@PlaceNumber{\llap{\normalfont
                \lst@numberstyle{\thelstnumber}\kern\lst@numbersep}\lst@linebgrd}\\%
     right:\def\lst@PlaceNumber{\rlap{\normalfont
                \kern\linewidth \kern\lst@numbersep
                \lst@numberstyle{\thelstnumber}}\lst@linebgrd}%
    }{\PackageError{Listings}{Numbers #1 unknown}\@ehc}}
\makeatother

\definecolor{dkgreen}{rgb}{0,0.6,0}
\definecolor{gray}{rgb}{0.5,0.5,0.5}
\definecolor{mauve}{rgb}{0.58,0,0.82}

\usepackage{graphicx}
\usepackage{tikz}
\usetikzlibrary{
    shapes.geometric, 
    arrows.meta, 
    positioning, 
    decorations.pathreplacing,
    shadows.blur,
    fit %
}

\renewbibmacro*{year}{%
  \iffieldundef{year}%
    {}%
    {\printfield{year}}%
}

\definecolor{sangria}{rgb}{0.57, 0.0, 0.04}
\definecolor{pinegreen}{rgb}{0.0, 0.47, 0.44}
\definecolor{rossocorsa}{rgb}{0.83, 0.0, 0.0}
\definecolor{ao}{rgb}{0.0, 0.0, 1.0}
\definecolor{deepjunglegreen}{rgb}{0.0, 0.29, 0.29}
\definecolor{dartmouthgreen}{rgb}{0.05, 0.5, 0.06}

\newcommand{\numtotalpypiscanned}{50,427\xspace}

\newcommand{\tool}{\textsc{PyFEX}\xspace}

\newcommand{\toolscan}{\textsc{PyFEXScan}\xspace}

\newcommand{\numnewmalware}{212\xspace}
\newcommand{\numalert}{222\xspace}

\newcommand{\numnewmalwaredownload}{91,648\xspace}

\newcommand{\numDoneorigin}{10,920\xspace}

\newcommand{\numDtwo}{417\xspace}

\def\7zip{\texttt{7zip}\xspace}

\def\e2fsck{\texttt{e2fsck}\xspace}

\def\mke2fs{\texttt{mke2fs}\xspace}

\DeclareAcronym{asan}{
    long={AddressSanitizer},
    short={ASan},
    cite={serebryany_addresssanitizer_2012}
}
\DeclareAcronym{ubsan}{
    long={UndefinedBehaviorSanitizer},
    short={UBSan},
    cite={undefinedbehaviorsanitizer_2013}
}
\DeclareAcronym{msan}{
    long={MemorySanitizer},
    short={MSan},
    cite={stepanov_memorysanitizer_2015}
}
\DeclareAcronym{put}{
    long={program under test},
    short={PUT},
}
\DeclareAcronym{ssrf}{
    long={Server-Side Request Forgery},
    short={SSRF},
}
\DeclareAcronym{xss}{
    long={Cross-Site Scripting},
    short={XSS},
}
\DeclareAcronym{toctou}{
    long={time-of-check to time-of-use},
    short={TOCTOU},
}
\DeclareAcronym{ebpf}{
    long={Extended Berkeley Packet Filter},
    short={eBPF},
    cite={ebpf}
}

\begin{document}

\title{\tool: Uncovering Evasive Python-based Threats via Resilient and Exhaustive Path Exploration}

\author{Meng Wang}
\affiliation{%
  \institution{CISPA Helmholtz Center for Information Security}
  \city{Saarbrücken}
  \country{Germany}}

\author{Yue Ma}
\affiliation{%
  \institution{University of Electronic Science and Technology}
  \city{Chengdu}
  \country{China}}

\author{Majid Garoosi}
\affiliation{%
  \institution{CISPA Helmholtz Center for Information Security}
  \city{Saarbrücken}
  \country{Germany}}

\author{Wenting Fan}
\affiliation{%
  \institution{Shanghai Jiao Tong University}
  \city{Shanghai}
  \country{China}}

\author{Liwei Guo}
\affiliation{%
  \institution{University of Electronic Science and Technology}
  \city{Chengdu}
  \country{China}}

\author{Jianqiang Wang}
\affiliation{%
  \institution{Max Planck Institute for Security and Privacy}
  \city{Bochum}
  \country{Germany}}

\author{Ali Abbasi}
\affiliation{%
  \institution{CISPA Helmholtz Center for Information Security}
  \city{Saarbrücken}
  \country{Germany}}

\begin{abstract}
The rapid expansion of the Python ecosystem has fueled two distinct but converging threats: adversaries increasingly target the software supply chain via the Python Package Index (PyPI), while also building evasive, cross-platform malicious binaries compiled from source code written in Python.
Current program analysis techniques struggle to address this dual threat. 
Static analysis based tools are often blinded by runtime obfuscation and compiled bytecode, while dynamic analysis based ones are fragile, prone to evasion by environmental guardrails, and often terminate prematurely due to unsatisfied dependencies. 

To overcome these limitations, we present \tool, a resilient forced-execution engine. 
\tool explores a program’s behavioral space systematically by forcing execution across all conditional branches to bypass evasion checks. 
To address the fragility of dynamic execution, it introduces a novel resilient crash recovery mechanism that creates synthetic objects to satisfy failed operations at runtime, allowing analysis to proceed past fatal errors, and employs path merging to mitigate path explosion. 
\tool further incorporates an automated entry identification mechanism that proactively discovers and invokes dormant functions, exposing malicious logic hidden within uncalled APIs.
To demonstrate the efficacy of this engine, we built \toolscan, a proof-of-concept malware detector built on top of \tool.
Evaluated against both known malicious PyPI packages and real-world compiled binaries, \tool exposes critical behaviors missed by the existing state-of-the-art tools. 
In a live deployment on PyPI, \toolscan discovered \numnewmalware previously unknown malicious packages accounting for over \numnewmalwaredownload downloads, underscoring the necessity of resilient, exhaustive analysis for securing the Python ecosystem.

\end{abstract}




\maketitle

\section{Introduction}
\label{sec:intro}

Python's emergence as the world's most popular programming language~\cite{tiobeTIOBEIndex} is a double-edged sword. 
While its simplicity and rich ecosystem have driven innovation across web infrastructure, machine learning, and scientific computing, its ubiquity has also made it a prime vehicle for adversaries to deploy what we collectively refer to as \textit{Python-based threats}. 
This threat landscape is twofold: first, attackers leverage Python to develop cross-platform malware~\cite{APT37, PythonBasedStealer}, often packaging compiled Python bytecode into standalone executables using tools like PyInstaller~\cite{pyinstaller}. 
This approach allows adversaries to leverage Python's rich ecosystem for rapid development while simultaneously evading signature-based defenses~\cite{malwareusepython}. 
Second, adversaries directly target the software supply chain by publishing malicious Python packages to the Python Package Index (PyPI). 
With PyPI receiving over 500 malware submissions monthly~\cite{pypiMalwarePackage}, the scale of this dual threat is substantial, indicating that Python’s ecosystem is increasingly targeted by attackers.

To counter these risks, a range of analysis techniques has been proposed~\cite{maloss, needle, malguard, malbindit, guarddog, Ossfpackageanalysis, ea4mp, CEREBRO, oscar} to detect these Python-based threats. Yet, despite significant research efforts, existing methods are still hindered by critical visibility gaps that leave malicious program capabilities within their blind spots. 

First, existing static analysis based techniques~\cite{malguard, malbindit, CEREBRO, ea4mp} are limited by their lack of runtime information.
They are blinded by common evasion techniques, such as dynamic code generation (e.g., eval), code encoding, or fetching payloads from remote servers~\cite{limitstaticml, ambiguoustranslation}. 
Even state-of-the-art tools like Malguard~\cite{malguard} acknowledge limitations against such runtime obfuscation. 
While theoretically possible to flag all obfuscated code, doing so causes unacceptably high false positives due to legitimate protection tools like PyArmor~\cite{githubGitHubDashingsoftpyarmor}. 
Furthermore, without runtime provenance, static systems often fail to distinguish between user-requested sensitive operations (e.g., process termination) and malicious abuse.

Second, dynamic analysis based techniques~\cite{Ossfpackageanalysis, FV8, malmax, oscar} suffer from fragility and incomplete coverage. 
Traditional dynamic analysis approaches~\cite{Ossfpackageanalysis, oscar} struggle to identify correct entry points in modular libraries and are easily evaded by anti-analysis environment checks that suppress malicious behavior.
To address these challenges, state-of-the-art forced-execution techniques (e.g., X-Force~\cite{xforce} for executable binaries, FV8~\cite{FV8} for JavaScript, and MalMax~\cite{malmax} for PHP) have been proposed.
However, these tools provide an incomplete analysis of the target program due to fragility and limited scope. 
Their analysis often terminates prematurely upon encountering unrecoverable states such as offline C2 (Command and Control) servers, preventing the discovery of downstream behaviors. 
Additionally, they do not proactively recognize and trigger implicit entry points, leaving dormant functions untouched by the analysis tool. 
Crucially, FV8~\cite{FV8} only applies forced execution to conditional blocks containing specific APIs (e.g., \texttt{eval()}) and relies on a deobfuscator to identify those conditional blocks. 
While this strategy mitigates path explosion, it fundamentally limits the analysis scope and consequently misses critical malicious behaviors. 
Similarly, MalMax~\cite{malmax} relies on a coarse-grained external PHP interpreter emulator, which limits both scalability and the granularity of collected information compared to inside-interpreter analysis.

Third, compiled Python bytecode represents a major blind spot. Existing Python package analyzers are overwhelmingly designed for source code and cannot process the compiled binaries (.pyc) increasingly used in real-world attacks~\cite{scworldPythonByte}. 
Approaches like FV8~\cite{FV8} are ill-suited for this domain, as their reliance on Abstract Syntax Tree (AST) instrumentation is fundamentally incompatible with compiled bytecode, where AST reconstruction is unreliable. 
Although decompilation is a potential workaround, attackers frequently exploit decompiler bugs to prevent successful source code recovery~\cite{pyfet}, rendering this approach ineffective.

Consequently, the visibility gaps inherent in existing analysis techniques inevitably degrade the efficacy of downstream detection systems. Whether relying on traditional machine learning classifiers~\cite{limitstaticml, malguard} or modern Large Language Models (LLMs)~\cite{malguard, SpiderScan}, these detectors are fundamentally constrained by the incompleteness of their input data. 
These limitations motivate the design of \tool, the first forced-execution engine specifically designed to analyze the full spectrum of Python-based threats, covering both source code and compiled bytecode. 
By integrating directly into the CPython interpreter~\cite{cpython}, \tool achieves fine-grained control over the runtime execution state, enabling novel capabilities designed to overcome the fragility and blindness of prior work.
Specifically, \tool systematically \textit{Forces Execution} down conditional branches to overcome evasive guardrails (e.g., environment checks) and employs a novel \textit{Resilient Crash Recovery} mechanism. 
This mechanism intercepts fatal runtime errors, such as connection attempts to abandoned C2 servers, and creates \textit{Synthetic Objects} to satisfy the operation, ensuring analysis proceeds deep into the execution chain where critical behaviors often reside. 
To ensure scalability without sacrificing semantic information, \tool utilizes a \textit{Semantic-Preserving Path Merging} strategy that consolidates divergent execution states while preserving the data provenance required to distinguish benign functionality from malicious intent. 
Finally, it incorporates a \textit{Dormant Function Analysis} module that actively identifies and invokes uncovered callable objects, exposing malicious logic hidden in non-standard entry points.

These capabilities enable \tool to resiliently expose runtime behaviors that attackers actively attempt to conceal. 
Crucially, the rich semantic information preserved during analysis allows for the accurate differentiation of benign packages with sensitive functionalities from genuinely malicious ones. 
To demonstrate the efficacy of \tool, we developed \toolscan, a proof-of-concept detector that integrates \tool's deep analytical traces with the reasoning capabilities of a Large Language Model (LLM)~\cite{llmnpmmalicious}.
Our evaluation against \numDoneorigin malicious PyPI packages and \numDtwo malicious executables collected in the wild demonstrates \toolscan's superior detection efficacy. 
Furthermore, in a 90-day live deployment analyzing daily PyPI uploads, \toolscan discovered \numnewmalware previously unknown malicious packages accounting for over \numnewmalwaredownload downloads, confirming its real-world utility.

\noindent
\textbf{Contributions.} Our approach has the following highlights:
\begin{compactitem} 
\item \textbf{Novel Analysis Framework.} We design and implement \tool, the first forced-execution based analysis tool integrated directly into the CPython interpreter, capable of analyzing both Python source code and compiled bytecode. This approach enables precise state manipulation and deep exploration of both source and binary-only threats.

\item \textbf{Resilient Execution.} We introduce a resilient crash recovery mechanism that utilizes on-demand object synthesis to prevent analysis termination during fatal runtime errors (e.g., dead C2 servers). Coupled with a Dormant Function Analysis strategy that actively identifies and invokes unexecuted callable objects, this design allows the engine to generate more complete behavioral traces than existing fragile dynamic analyzers.

\item \textbf{Semantic-Aware State Management.} We build a scalable path-merging method that consolidates divergent execution paths at post-dominators while preserving the data provenance. 
This semantic richness enhances the performance of downstream detectors, enabling them to distinguish between user-intended sensitive actions and malicious ones.

\item \textbf{Empirical Validation and Discovery.} We evaluate \toolscan on a rigorously calibrated dataset, where we identified and corrected 112 misclassified benign packages in established ground-truth benchmarks used by other academic works~\cite{malguard, ea4mp}. Furthermore, our live deployment discovered \numnewmalware previously unknown malicious packages having \numnewmalwaredownload times of downloading on PyPI, underscoring the necessity of resilient dynamic analysis.

\end{compactitem}

\section{Background}
\label{sec:background}
\subsection{CPython Interpreter}
CPython~\cite{cpython} serves as the reference implementation and dominant interpreter for the Python ecosystem. We architected \tool directly on top of CPython to ensure broad applicability to real-world threats. 
The interpreter operates by compiling source code into bytecode instructions, which are then executed by the stack-based Python Virtual Machine. 
This architecture reduces complex high-level logic to a finite set of predictable operations, providing ideal instrumentation points for fine-grained analysis.

By instrumenting the execution loop, \tool gains direct control over control-flow instructions (e.g., \texttt{POP\_JUMP\_IF\_FALSE}). 
This allows the engine to override boolean checks and force execution down evasive paths that malware authors attempt to hide. 
Furthermore, this deep integration also enables \tool to intercept fatal runtime exceptions such as connection errors to offline C2 servers at the instruction level. 
By synthesizing counterfactual states to satisfy these operations, execution can continue, exposing the logic that would otherwise remain unseen. 
Operating inside the interpreter gives us the ability not only to observe but also to control, transforming a simple runtime into a rich analysis platform.

\subsection{Python-Based Malware}
In this paper, we call Python-based threats as malicious software developed using the Python language. 
This threat contains two vectors: supply chain infiltration via public repositories and standalone executables compiled from Python scripts using popular, open-source tools such as PyInstaller~\cite{pyinstaller} and Py2Exe~\cite{py2exe} to evade traditional defenses~\cite{malwareusepython}.

Adversaries target the Python Package Index (PyPI) to distribute malware directly into developer environments. Techniques such as typosquatting (e.g., colourama vs. colorama~\cite{Colorama}) and dependency confusion exploit PyPI's open model and the automatic resolution of transitive dependencies, allowing malicious code to propagate silently across the ecosystem.

Attackers~\cite{bleepingcomputerUkrainesArmy,thehackernewsStealerMalware} are increasingly using tools like PyInstaller~\cite{pyinstaller} or Py2Exe~\cite{py2exe} to bundle malicious Python scripts into a single executable. 
For an attacker, this process offers a powerful twofold advantage: simplified deployment and effective evasion. 
The technique serves as a formidable layer of obfuscation against security tools.
When a standard static analyzer scans the compiled executable, it does not see Python bytecode; instead, it sees the legitimate bootloader of the packaging tool. 
The actual malicious script is typically compressed and archived within the executable's data section, remaining invisible to scanners that are not specifically designed to unpack these formats. 
Runtime analysis is similarly challenged. Upon execution, the bootloader first unpacks the interpreter and the malicious bytecode into a temporary directory before executing it, often in a separate process. 
This means the malicious behavior is detached from the initial file that a sandbox might be monitoring. 
This method is highly effective and has been used to deliver a wide range of malware, from information stealers~\cite{AkiraStealer} to the recent wave of Python-based ransomware~\cite{PyLocky}, posing a significant challenge to detection.

These vectors present distinct challenges: supply chain attacks primarily involve source code, while standalone executables conceal logic within compiled bytecode. 
Despite the prevalence of compiled threats, existing research has largely focused on source-based PyPI analysis. 
\tool addresses this critical blind spot by providing the first unified platform capable of analyzing both Python source code and compiled binaries

\section{Motivation Example: PyLocky Ransomware}
\label{sec:motivation}
To illustrate the limitations of existing tools and motivate the design of \tool, we examine PyLocky, a real-world ransomware that exemplifies the complexity of current attackers and serves as a motivation example. 
PyLocky is developed in Python and packaged into a standalone executable using PyInstaller~\cite{pyinstaller}, a legitimate tool often abused by attackers to obscure malicious code. 
This packaging, combined with runtime evasions,  renders PyLocky particularly difficult for existing tools to detect and analyze.

\begin{listing}[!ht]
\begin{minted}
[
frame=lines,
framesep=2mm,
baselinestretch=1.2,
fontsize=\scriptsize,
linenos,
numbersep=1pt %
]
{python}
LockRAM = int(round(system_ram))
#evasive technique for sandbox detection
if LockRAM < 4:
    time.sleep(9999)
# command server
url = "http://centredentairenantes.fr/wp-system.php" 
r = requests.get(url)
....
# later code for encrypting files
\end{minted}
\caption{Code Snippet of PyLocky Ransomware (Decompiled and Deobfuscated)}
\label{listing:motivation_example}
\end{listing}

Listing~\ref{listing:motivation_example} shows a simplified representation of PyLocky's core logic. The malware was distributed via spam email campaigns targeting European businesses~\cite{PyLocky}. Its execution involves several stages designed to thwart analysis before the final encryption payload is deployed.

\subsection{Challenge 1: Packaging and Obfuscation}
PyLocky's primary delivery mechanism is a PyInstaller executable. 
This immediately presents a barrier to the existing malware analysis tools, including state-of-the-art static analysis systems like Malguard~\cite{malguard} or hybrid analysis systems like OSSF Package Analysis~\cite{Ossfpackageanalysis}, which are designed for Python source code (.py) files. 
These tools cannot directly parse the compiled bytecode (.pyc) embedded within the executable. While decompilation is theoretically possible, it is often unreliable and can be easily defeated by anti-analysis techniques~\cite{pyfet, pylingual}. 
Furthermore, runtime decoding often hides the final payload, presents a further barrier to any form of static inspection, as admitted by Malguard~\cite{malguard}.

\noindent\textbf{\tool's solution}: By operating directly within the CPython interpreter, \tool executes and analyzes the bytecode, bypassing the need for unreliable decompilation and directly observing the code as it runs from an internal perspective, regardless of initial packaging.

\subsection{Challenge 2: Runtime Evasion}
PyLocky employs explicit anti-sandbox checks. One of them is the RAM check (Line 3) in Listing~\ref{listing:motivation_example}): if the system has less than 4GB of memory, it enters the sleep status without doing any malicious behaviors. 
Conventional dynamic analysis sandboxes typically run with minimal resources for scalability and would trigger this sleep condition. 
The analysis would time out long before the malware performs any encryption or C2 communication, resulting in a false negative report.

\noindent\textbf{\tool's solution}: \tool's forced execution mechanism instruments bytecode-level conditional jumps. It will discover both branches regardless of the conditions, ensuring the later malicious code including subsequent C2 communication and encryption logic are always exposed during analysis.

\subsection{Challenge 3: Fragility}
PyLocky relies on communication with a C2 server (Line 7) to potentially retrieve configuration or commands. 
If this server is offline during analysis (a common scenario for older malware samples or post-attack analysis), the network request will fail. 
Existing forced-execution tools like MalMax~\cite{malmax} or FV8~\cite{FV8}, if ported, would crash or terminate the analysis of this path upon encountering the unhandled network exception. 
This prevents observation of the critical downstream encrypt\_files routine. 
Furthermore, FV8's AST-based approach is fundamentally incompatible with bytecode analysis. 
Naively expanding FV8's scope would also be computationally intractable due to path explosion.

\noindent\textbf{\tool's solution}: \tool has a resilient crash recovery mechanism to intercept the exceptions. Instead of crashing, it synthesizes a counter-factual state and allows execution to continue seamlessly into the later encrypt\_files function. 
Combined with path merging to manage complexity, \tool ensures that the complete encryption behavior is revealed, regardless of external dependencies or evasive checks.

This motivating example illustrates the need for an analysis system that can systematically explore the execution space to reveal the target program's hidden malicious behaviors, both in binary format and obfuscated code.

\section{Design}
\label{sec:design}

\begin{figure*}[ht!]
    \centering
    \includegraphics[width=\textwidth]{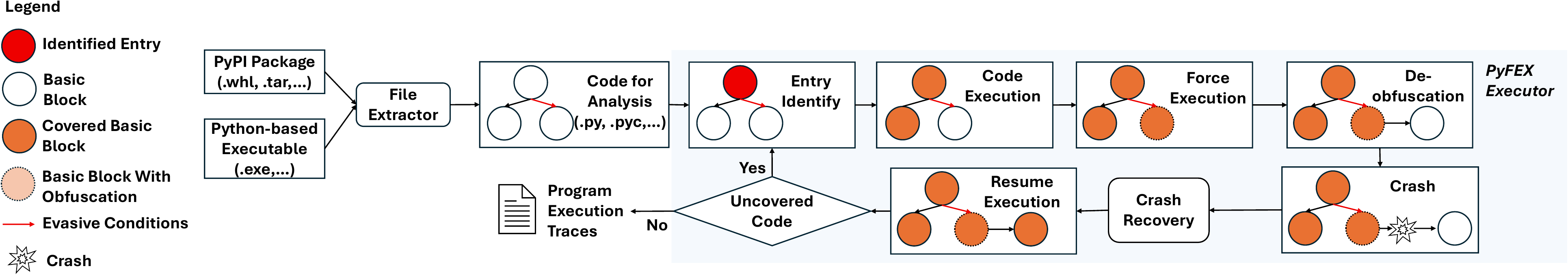}
    \caption{Overview of \tool}
    \label{fig:overview}
\end{figure*}

This section details the design of \tool, a dynamic analysis engine built to expose the behaviors of evasive Python-based threats. Malware developers increasingly abuse Python’s dynamic features, such as runtime metaprogramming and environmental checks, to conceal malicious logic from static analysis and evade dynamic analysis by placing malicious behavior behind environmental checks. 
The ultimate goal of these techniques is to ensure that critical payloads remain unobserved during analysis.

To address these challenges, as shown in Fig.~\ref{fig:overview}, \tool systematically explores the execution space of the target program by four complementary designs to ensure robustness and scalability. 
First, we employ \textbf{Context-Aware Forced Execution} to explore every conditional path. Unlike prior work, \tool uses a novel state-sharing model that synchronizes objects within the same closure across branches, ensuring forced paths inherit the valid computational state required to proceed without crashing. 
Second, to address the fragility of dynamic analysis, we introduce \textbf{Resilient Crash Recovery}. When fatal errors occur, the engine intercepts the exception and synthesizes a \textit{Synthetic Object} to satisfy the operation. 
These objects actively record the crash origin and track their propagation lineage, allowing the analysis to reconstruct the semantic context of hidden behaviors even after failures.
To manage the complexity of exhaustive exploration, \tool implements \textbf{Semantic-Preserving Path Merging}, which consolidates divergent execution paths at their post-dominator points while rigorously retaining the data provenance of sensitive objects. 
Finally, to ensure complete coverage, we incorporate a \textbf{Dormant Function Analysis} phase. If callable objects remain uninvoked after the initial trace, \tool proactively executes them using synthesized inputs, exposing logic hidden behind non-standard entry points. 
The output of \tool is a set of comprehensive execution traces that capture not only the sequence of operations but also the semantic context in which they occur, providing a robust foundation for malware detection and forensic analysis.

\subsection{Forced Execution}
\label{sec:meth:forced_execution}
A fundamental challenge in dynamic malware analysis is achieving sufficient code coverage against evasion conditions that check for environmental artifacts (e.g., debuggers, sandboxes) and suppress malicious payloads if detected. 
Conventional dynamic analysis~\cite{Ossfpackageanalysis}, which follows a single concrete execution path, typically fails to satisfy these predicates. 
As a result, the malicious logic is never executed, the visibility gap remains, and the malware is misclassified as benign.

\begin{table}[ht!]
    \scriptsize
    \centering
    \caption{Sample Instrumented Bytecodes for Counter-Factual Execution}
    \label{tab:opcode_force}
    \begin{tabular}{p{0.4\linewidth} p{0.5\linewidth}}
        \toprule
        \textbf{Bytecode Instr.} & \textbf{ Action} \\
        \midrule

        \multicolumn{2}{l}{\textit{\textbf{1. Conditional Branching}}} \\
        \texttt{POP\_JUMP\_IF\_FALSE} & Invert the condition \\
        \texttt{POP\_JUMP\_IF\_TRUE}  & Invert the condition \\
        \texttt{JUMP\_ABSOLUTE}       & Redirect to another code path. \\
        \midrule

        \multicolumn{2}{l}{\textit{\textbf{2. Looping Constructs}}} \\
        \texttt{FOR\_ITER}           & Force premature loop termination/Inject items to prolong iteration. \\
        \texttt{JUMP\_BACKWARD}      &  Force an early exit from the loop. \\
        \midrule

        \multicolumn{2}{l}{\textit{\textbf{3. Exception Handling}}} \\
        \texttt{SETUP\_FINALLY}       & Inject a synthetic exception  \\
        \texttt{RAISE\_VARARGS}      & Suppress the exception\\
        \texttt{JUMP\_IF\_NOT\_EXC\_MATCH} & Force a match to enter a specific \texttt{except} \\
        \midrule
        
        \multicolumn{2}{l}{\textit{\textbf{4. Context Management}}} \\
        \texttt{SETUP\_WITH}          & Simulate an exception \\
        \texttt{WITH\_EXCEPT\_START}  & Force the \texttt{\_\_exit\_\_} method to suppress an exception \\
        \midrule

        \multicolumn{2}{l}{\textit{\textbf{5. Generators \& Asynchronous Flow}}} \\
        \texttt{YIELD\_VALUE}         & Substitute the yielded value/Inject \texttt{StopIteration} to terminate. \\
        \texttt{GET\_AWAITABLE}       & Replace the awaitable with a mock that resolves immediately with a counter-factual result. \\

        \bottomrule
    \end{tabular}
\end{table}

To overcome this, \tool implements a forced execution engine that explores the program's behavioral space by traversing all control flow paths. 
Unlike prior approaches that rely on source-to-source transformation or Abstract Syntax Tree (AST) rewriting (e.g., FV8~\cite{FV8}), \tool instruments the CPython interpreter directly at the bytecode level. 
This design choice is critical for robustness: while AST-based instrumentation is fragile and easily broken by anti-parser techniques or syntax errors~\cite{pyfet, pylingual}, bytecode instrumentation operates on the definitive, low-level instructions executed by the Python Virtual Machine (PVM). 
This renders our analysis opaque to the malware and compatible with obfuscated or compiled code where AST reconstruction is unreliable.

Specifically, we modified the C-level bytecode evaluation loop to intercept control-flow-altering bytecodes (e.g., \texttt{POP\_JUMP\_IF\_FALSE}, \texttt{FOR\_ITER} in Table~\ref{tab:opcode_force}). When \tool encounters a control flow related bytecode (full bytecodes list in Table~\ref{tab:pyfex-bytecodes-310}), it forks the execution state. 
The original process continues along the concrete path dictated by the runtime values, while a forked counter-factual process is forced down the alternative branch by manipulating the instruction pointer to the alternative jump target. 
This principle is applied recursively to \texttt{if/else} blocks, loops, and exception handlers, ensuring that the analysis proceeds past guards that would otherwise halt execution.

To illustrate the rationale of our design, we show how \tool handles two distinct, real-world malware families: Pbot~\cite{Pbot}, known for its bytecode format and multi-layered obfuscation, and NodeStealer~\cite{NodeStealer}, which employs direct anti-analysis checks.

\begin{listing}[!ht]
\begin{minted}
[
frame=lines,
framesep=2mm,
baselinestretch=1.2,
fontsize=\scriptsize,
linenos,
numbersep=1pt %
]
{python}
...
butchertemper="4"
preventpennyworths="m"
timesmammet="txlbmStlQZ+uNYoFoVz8yrkEI1Um7OkJHXE"
villainweapon=exec
...
villainweapon('mannerseastern='+placentiocherish)
...
villainweapon('maidenheadgoing='+graciousstory)
...
\end{minted}
\caption{Code Snippet Pbot Malware}
\label{listing:Pbot_obfuscated}
\end{listing}

\textbf{Defeating Multi-Layered Obfuscation}: This technique is equally effective against multi-layered obfuscation, as seen in the Pbot~\cite{Pbot} malware as shown in Listing~\ref{listing:Pbot_obfuscated}. 
Static analysis tools were easily defeated by its bytecode format as it abuses the bugs in the existing decompilers~\cite{pyfet}. 
Even if the de-compilation succeed, static analysis based malware detector can easily flag this package as suspicious while multi-layers of encoding and dynamic execution via eval() still make the analysis tool hard to answer what exactly are the malicious behaviors or indicator of compromise. 
While a conventional dynamic analysis might deobfuscate the first layer, it remains blind to any conditional logic within the deobfuscated payload. 
This is a critical flaw, as sophisticated malware often contains further evasive checks in its later stages. 
\tool's approach is more resilient. 
\tool executes the code to naturally bypass the initial de-obfuscation layers. 
Crucially, as the payload is materialized and executed by the interpreter, \tool recursively applies forced execution to the newly generated bytecode. 
This ensures that even conditional logic nested within dynamically generated payloads is fully explored.

\begin{listing}[!ht]
\begin{minted}
[
frame=lines,
framesep=2mm,
baselinestretch=1.2,
fontsize=\scriptsize,
linenos,
numbersep=2pt %
]
{python}
blacklistedProcesses = [
    "ollydbg", "vmwareuser", "vgauthservice",...] 
#multiple processes indicating VMs or debuggers
def check_process() -> bool:
    for proc in psutil.process_iter():
        process_name = proc.name().lower()
        for blacklisted_name in blacklistedProcesses:
            if blacklisted_name in process_name:
                    return false
    return true
if check_process():
    do_malicious()
else:
    do_benign()
\end{minted}
\caption{Anti-Analyzer: NodeStealer}
\label{listing:nodestaler}
\end{listing}

\textbf{Bypassing Anti-Analysis Evasion}: Consider the NodeStealer~\cite{NodeStealer} malware shown in Listing~\ref{listing:nodestaler}, which explicitly checks for the presence of debugger-related processes and terminates if any are found. 
Attempting to create a perfect sandbox environment that is indistinguishable from a real user's machine is an intractable arms race. forced execution fundamentally sidesteps this problem. 

\begin{figure}[ht!]
    \centering
    \includegraphics[width=0.45\textwidth]{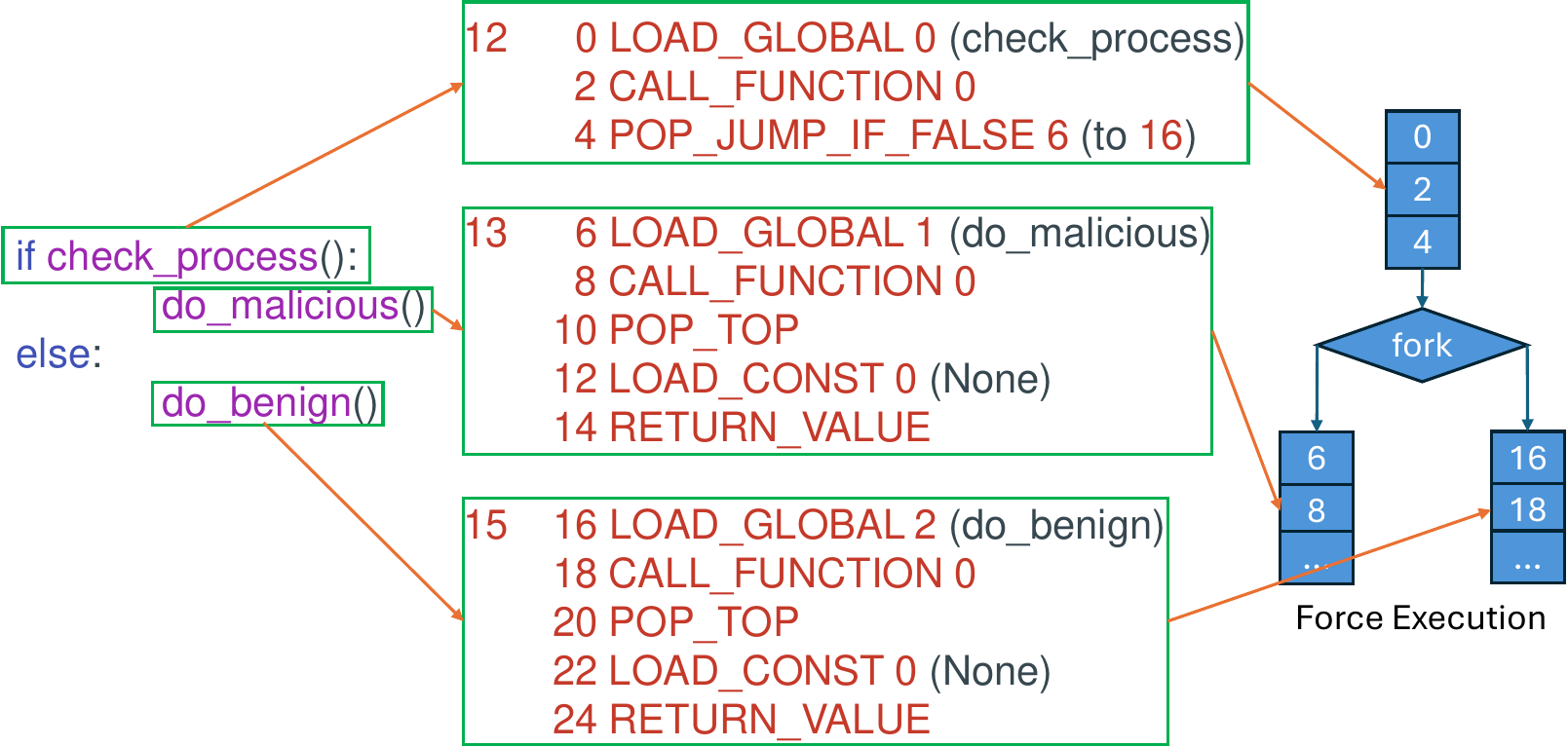}
    \caption{Nodestealer Forced Execution of One Conditional Statement}
    \label{fig:forceexecution}
\end{figure}

Upon reaching the conditional check for a debugger, \tool does not attempt to satisfy the predicate. 
Instead, as shown in Figure~\ref{fig:forceexecution}, it spawns a new process to discover the other execution path. One path follows the concrete outcome (the return false call), while a second, counter-factual path is created that forces execution down the alternative branch. 
This approach guarantees that the malicious payload protected by an anti-analysis technique is reached and analyzed, regardless of the analysis environment's characteristics.

\subsection{Resilient Crash Recovery} 
\label{sec:meth:crash_prevention}

A major limitation of existing forced-execution systems is fragility: analysis often terminates prematurely when a forced path encounters a fatal error, such as a network request to an offline C2 server. This leads to the invisibility of the downstream code.
To address this, \tool introduces a Resilient Crash Recovery mechanism that synthesizes valid execution states to recover from fatal runtime errors in incomplete or hostile environments.

\subsubsection{Bytecode Level Crash Recovery }
\begin{table}[ht!]
    \scriptsize
    \centering
    \caption{Part of Python Bytecode Instructions and Potential Terminating Exceptions}
    \label{tab:crash_prevention}
    \begin{tabular}{ll}
        \hline \\[-1.5ex]
        \textbf{Bytecode Instr.} & \textbf{Potential Exception} \\
        \hline \\[-1.5ex]

        \multicolumn{2}{l}{\textit{Attribute and Name Access}} \\
        \texttt{LOAD\_ATTR} & Attribute does not exist\\
        \texttt{LOAD\_GLOBAL} &  Global variable is not defined \\
        \texttt{LOAD\_FAST} & Variable is used before assignment \\
        \hline \\[-1.5ex]

        \multicolumn{2}{l}{\textit{Container Operations}} \\
        \texttt{BINARY\_SUBSCR} & Index is invalid \\
        \texttt{UNPACK\_SEQUENCE} & Object fails to unpack \\
        \hline \\[-1.5ex]

        \multicolumn{2}{l}{\textit{Arithmetic Operations}} \\
        \texttt{BINARY\_FLOOR\_DIVIDE} & Division by zero \\
        \texttt{BINARY\_MODULO} & Modulo by zero \\
        \hline \\[-1.5ex]

        \multicolumn{2}{l}{\textit{Function and Control Flow}} \\
        \texttt{CALL\_FUNCTION} & Object is not callable \\
        \texttt{FOR\_ITER} & Object is not iterable \\
        \hline
    \end{tabular}
\end{table}

Our approach is built upon a key architectural insight into interpreted languages: exception handling is centralized within specific bytecode instructions. 
While a program may crash for various reasons at the source-code level, these failures invariably manifest as exceptions raised by a finite set of bytecode instructions. 
Table~\ref{tab:crash_prevention} shows a subset of the instructions we instrument and the exceptions they commonly trigger (full version is in Table~\ref{tab:pyfex-bytecodes-310}). 
This observation allows us to construct a robust recovery system by instrumenting this finite set of instructions, avoiding the intractable task of patching every possible high-level error scenario.

For instance, a single instrumentation point on the bytecode \texttt{CALL\_FUNCTION} can handle failures caused by missing functions, signature mismatches, or non-callable objects. 
Rather than addressing these errors individually, the \tool interpreter intercepts the exception, clears the error state, and returns a \textit{Synthetic Object} to satisfy the operation as the return object of the function call. 
This ensures that execution proceeds seamlessly past the point of failure to further analyze the downstream code.

Crucially, a \textit{Synthetic Object} is not only a static placeholder. 
It acts as a dynamic proxy capable of accepting any operation. 
Whether the malware attempts to access an attribute, invoke a method, or iterate over the object, the proxy recursively returns a new \textit{Synthetic Object}. 
This mechanism resolves common crash scenarios: accessing an index out of bounds (e.g., \texttt{my\_list[10]}) returns a generic \textit{Synthetic Object} instead of raising an \texttt{IndexError}; querying a missing dictionary key (e.g., \texttt{my\_dict['config']}) suppresses the \texttt{KeyError}; and failed imports trigger the creation of a synthetic module populated with synthetic functions. 
Consequently, \tool enables the exploration of downstream logic that depends on these otherwise unavailable resources.

\subsubsection{Semantic Reconstruction via Trace Properties}
While the crash-recovery mechanism enables \tool to uncover previously unreachable code, only execution is insufficient. 
We must also recover meaningful semantic information from these execution paths. 
Performing computations on simple empty placeholders yields no insight. 
To address this, \tool implements a data provenance tracking system.

Our core design principle is that any bytecode instruction consuming a \textit{Synthetic Object} produces a new \textit{Synthetic Object} as output. 
Crucially, this output is not a blank slate. Each \textit{Synthetic Object} maintains a \textit{Trace Property} which serves as a symbolic expression that records the object's origin and the history of operations performed on it. 
Instead of performing meaningless arithmetic on arbitrary values, \tool symbolically records the computation. 
For instance, if a \textit{Synthetic Object} is used in an \texttt{ADD} instruction with the integer 1, the result is a new \textit{Synthetic Object} that internally records its lineage as \texttt{ADD(Synthetic\_Object, 1)} which will be stored in its trace property. 
If this new object is then used in a \texttt{BINARY\_SUBSCR} operation, the final Synthetic Object's trace property would become \texttt{SUBSCR(ADD(Synthetic\_Object, 1), 2)}
This ensures that even when concrete data is unavailable due to a crash, the system will still be able to recover some of the semantic context of the data flow.

\begin{figure*}[ht!]
    \centering
    \includegraphics[width=0.9\textwidth]{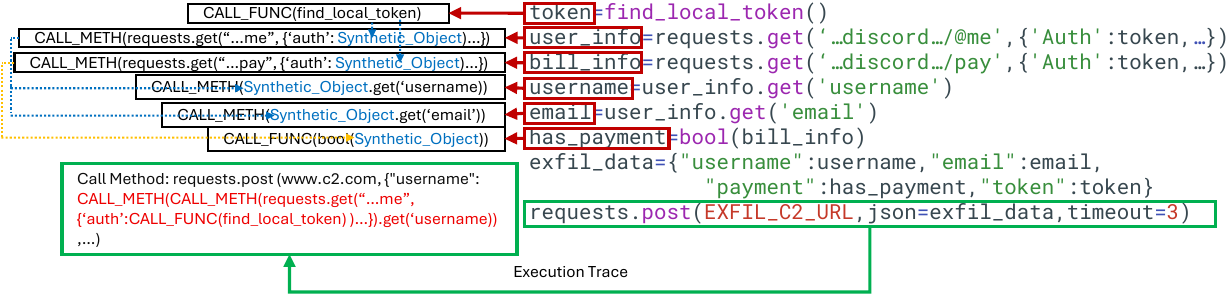}
    \caption{VBucks Malware: Trace Property}
    \label{fig:vbuckstraceproperty}
\end{figure*}

This mechanism is critical for analyzing real-world threats like the VBucks~\cite{vbucksmalware}, a malware family that targets stealing sensitive information from the victim's machine.
As shown in part of the simplified, de-obfuscated, and decompiled code in Figure~\ref{fig:vbuckstraceproperty}, the malware first attempts to find a Discord token on the victim's machine using a function like \texttt{find\_local\_token}. 
In a naive sandbox environment, this function will fail to find the required token path, and the program will terminate, leaving its true malicious behavior unobserved.
\tool overcomes this challenge through a combination of forced execution and crash prevention.
We forcefully execute the code path that assumes a token is found. 
When the code attempts to read the non-existent token file, our crash prevention mechanism intercepts the failure and synthesizes a Synthetic Object for the token variable. The analysis then proceeds to the subsequent \texttt{requests.get} calls to the Discord API. These calls will naturally fail because they use an invalid token which is a \textit{Synthetic Object}. 
However, because the \texttt{requests.get} function is consuming a Synthetic Object, \tool again intervenes. It bypasses the crash and creates a new Synthetic Object for the \texttt{user\_info} variable. 
This new object's trace property now records the entire chain of events, as illustrated in Figure~\ref{fig:vbuckstraceproperty}.
Similarly, as this \texttt{user\_info} object is used to derive other variables, such as \texttt{username} and \texttt{email}, the trace property is continuously propagated. 
When the malware makes its final \texttt{requests.post} call to the attacker's server, the data being sent contains Synthetic Objects whose trace properties can be recursively resolved. 
The final logs do not just show a network call; they reveal a complete, semantic narrative: data originating from an attempt to read a Discord token file was processed by several functions and ultimately sent to an external server.

By synthesizing objects to satisfy otherwise-crashing operations, \tool uncovers logic contingent on data from unavailable sources. 
Simultaneously, the propagation of Synthetic Objects with detailed provenance tracking preserves a rich semantic trace of all computations that would have occurred, providing a complete view of the malware's capabilities, which would be used in further analysis.

\subsection{Semantic-Preserving Path Merging}

While forced execution and crash prevention are powerful for exposing otherwise hidden code, they introduce two major challenges: the exponential growth of execution paths (path explosion) and the potential for semantic incoherence when forcing execution down infeasible paths. 
Existing forced-execution systems such as FV8~\cite{FV8} and MalMax~\cite{malmax} mitigate path explosion by significantly limiting their analysis scope or simply terminating upon encountering a crash. 
While these design choices reduce the search space, they fundamentally sacrifice the completeness of exposed behaviors, a compromise we have shown to be unsuitable for the diverse and evasive Python-based malware ecosystem (as detailed in Section~\ref{sec:eval:deep_analysis}). 
This subsection details \tool's approach to managing these complexities, which aims to maximize path exploration while rigorously preserving program semantics.

\subsubsection{Integrated Path and State Merging}
Forced execution, by its nature, creates divergent execution processes at every conditional branch. 
For nested conditions, this leads to an exponential increase in active paths, a formidable challenge known as path explosion. 
This issue makes comprehensive analysis computationally infeasible. 
To ensure scalability while preserving completeness, \tool implements a light-weight post-dominator-based path merging strategy for the forced execution.

Our strategy begins by allowing divergent execution processes, each representing a distinct path through a conditional branch, to run independently until they reach their immediate post-dominator. 
For each \texttt{PyCodeObject}, \tool computes this merge point on an intra-procedural control-flow graph built from bytecode offsets, explicit jump targets, fall-through edges, and exception-handler metadata, which makes CFG construction straightforward.
This defines the unique point in the control flow graph where all divergent paths from a specific branch must eventually reconverge.
When the first process arrives, its corresponding post-dominator continues execution. Once its slower peers arrive at the post-dominator later, \tool initiates the merging process. 
The merging process is designed to preserve the full spectrum of observed behaviors and prune the slower processes. 
It is noteworthy that the slower process does not simply disappear while its entire state, including local and global variables, imported modules, and other runtime objects, is logically transferred and associated with the continuing process. 
This means the surviving process now carries a superset of the state from both original paths.

During subsequent execution, when the merged process needs to use a variable or object, it employs a conflict resolution mechanism as follows: The process first attempts to use its own value for a computation. 
However, if its own value is a Synthetic Object (synthesized due to a previous crash) while the corresponding value from the pruned path is a concrete object, \tool prioritizes the concrete value. 
This ensures that valid data is used whenever possible. 
If the initial computation attempt with a preferred value results in a crash or error, and the merged state contains alternative values for the same variable (from the pruned path), \tool will systematically try each alternative value until the first one resolves the issue. 
This iterative attempt ensures that the analysis proceeds with a value that permits valid execution, revealing downstream logic that might depend on specific conditions met in a particular branch. 
If none of the available values (from either the original or the merged path) can satisfy the requirements of the bytecode instruction without error, \tool defaults to creating a new Synthetic Object, as described in the previous section on crash prevention. 
This ensures that analysis never halts, even in ambiguous or fully corrupted states.

\begin{listing}[!ht]
\begin{minted}
[
frame=lines,
framesep=2mm,
baselinestretch=1.2,
fontsize=\scriptsize,
linenos,
numbersep=1pt %
]
{python}
def check_vm_sanbox():
    if OS_TYPE == "Windows":
        check_windows_indicators()
    else if OS_TYPE == "Darwin":
        check_mac_indicators()
    else:
        check_linux_indicators()
def install_persistence():
    if OS_TYPE == "Windows":
        install_persistence_win()
    else if OS_TYPE == "Darwin":
        install_persistence_mac()
    else:
        install_persistence_linux()
def get_data():
    if OS_TYPE == "Windows":
        get_data_win()
    else if OS_TYPE == "Darwin":
        get_data_mac()
    else:
        get_data_linux()
check_vm_sanbox()
install_persistence()
get_data()
\end{minted}
\caption{Code Snippet of Xillenstealer Malware}
\label{listing:Xillenstealer}
\end{listing}

Listing~\ref{listing:Xillenstealer} presents the modified and simplified code snippet from an information stealer malware named XillenStealer~\cite{xillenstealer} for presentation purposes. It leverages Python's cross-platform capabilities to implement distinct logic for Windows, macOS, and Linux. 
For naive forced execution, this structure poses a significant challenge: a sequence of just three platform-dependent checks results in an exponential explosion of execution states ($3^3$ = 27 paths), rendering exhaustive analysis computationally intractable.

\tool addresses this via post-dominator merging. Upon reaching the convergence point of a conditional block (e.g., the completion of \texttt{check\_vm\_sandbox}), \tool merges the divergent paths into a single process. 
Crucially, while this merged process retains the behavioral traces from all branches (recording that Windows, macOS, and Linux indicators were all checked), it must adopt a single concrete state to proceed. 
\tool resolves this by prioritizing the concrete objects and objects not causing crashes, which in this scenario, the Linux branch is preferred as it matches the analysis environment. 
By prioritizing valid objects from the natural path, the system minimizes crashes in downstream code and ensures that subsequent operations remain compatible with the underlying OS. 
Consequently, the final report successfully captures the malware's cross-platform capabilities without incurring the resource cost of maintaining parallel processes

Furthermore, \tool's \textit{Resilient Crash Recovery} mechanism is vital to this unification process. 
When analyzing Windows-specific branches within a Linux environment, calls to unavailable APIs (e.g., winreg) would typically trigger fatal errors. 
\tool intercepts these failures and generates \textit{Synthetic Object}s to sustain execution. 
Although these objects act as placeholders, their associated provenance traces faithfully record that install\_persistence\_win() was invoked and attempted to access specific Windows APIs.
This data persistence is crucial during merging: even if the Windows process is pruned in favor of the Linux path, its provenance data captured within the shared or merged variables is preserved. 
This allows \tool to report that persistence mechanisms for Windows, macOS, and Linux were all identified, detailing the specific API calls for each platform despite the incomplete analysis environment.

\subsubsection{Scope-Aware Cross-Path State Sharing}
\label{sec:meth:state_sharing}
While post-dominator merging efficiently consolidates execution paths, a challenge arises when one process reaches a post-dominator significantly faster than its peer. 
The faster process might require an object that has only been resolved in the slower, still-running process. 
To address this, \tool implements a \textit{Scope-Aware State Sharing mechanism}. Unlike prior systems like MalMax~\cite{malmax} that restrict sharing to global artifacts, \tool extends sharing to any variables within the same closure scope, which is essential for analyzing Python malware that passes critical state between local scopes. 

The sharing mechanism is implemented via a shared runtime state manager accessible to all forked processes. 
When an active process encounters errors like an \texttt{Unavailable Objects} or attempts to consume a \textit{synthetic object} during a critical operation, it triggers a state query. 
This query inspects the execution frames of concurrently running peer processes. 
If a peer process holds a valid definition for the requested symbol within a compatible scope, the value is copied and injected into the requesting process. 
This allows paths to borrow the state being computed in parallel, bridging the gap between divergent execution timelines.

To maintain semantic integrity during this exchange, \tool applies a conflict resolution policy as follows:
The first rule is a preference for concreteness over \textit{Synthetic Objects}: if an active context holds a \textit{Synthetic Object} while another context has a concrete, initialized value, the concrete value is preferred. This helps the analysis proceed with real data rather than symbolic placeholders.
The second rule is that if multiple peers offer conflicting concrete values, \tool prioritizes the value derived from the natural (non-forced) execution path.
In such cases, \tool prioritizes the object from the context that followed the natural execution path, assuming it is more likely to be semantically valid for continued exploration, whereas forced paths may contain counter-factual data.

\begin{listing}[!ht]
\begin{minted}
[
frame=lines,
framesep=2mm,
baselinestretch=1.2,
fontsize=\scriptsize,
linenos,
numbersep=2pt %
]
{python}
for share in possible_shares:
    hd = test_connection(share)
    ...
    win32wnet.WNetAddConnection2(nr, None, None, 0)
...
ransomware_encrypt(share)
\end{minted}
\caption{Code Snippet of Fscoiety Locker Ransomware}
\label{listing:FscoietyLocker}
\end{listing}

The code in Listing~\ref{listing:FscoietyLocker}, adapted from the Fsociety Locker ransomware~\cite{FsocietyLocker}, first attempts to connect to all possible network devices before encrypting files on the accessible drives. 
To ensure full coverage, \tool forks its analysis at the loop. One "slower" process enters the loop to analyze the network connection and information collection logic, while a "faster" process bypasses it to directly analyze the ransomware\_encrypt function on Line 7. 
This creates a semantic challenge: the faster process reaches the encryption call, but its shared variable remains unresolved, as its value is only determined within the loop. Awaiting a full post-dominator merge would be inefficient. Instead, \tool employs its state-sharing mechanism. 
The faster process, detecting its unresolved shared object, queries its slower, concurrently-running peer and borrows the concrete shared value that has been resolved within the loop. 
This dynamic sharing allows the faster path to meaningfully analyze the ransomware\_encrypt function with a valid target, thus allowing the discovery of the ransomware's core logic, such as identifying which file extensions it targets for encryption.

\subsection{Dormant Function Analysis}
\begin{listing}[!ht]
\begin{minted}
[
frame=lines,
framesep=2mm,
baselinestretch=1.2,
fontsize=\scriptsize,
linenos,
numbersep=2pt %
]
{python}
class HTTPError(RequestException):
    # An HTTP error occurred
    __import__('\x62\x75\x69\x6c...').exec(...)
    ...
\end{minted}
\caption{Code Snippet of Ultrarequest}
\label{listing:ultrarequest}
\end{listing}

Identifying correct entry points is a critical yet challenging prerequisite for dynamic analysis, particularly for malicious libraries like ultrarequest~\cite{ultrarequst} (Listing~\ref{listing:ultrarequest}) that conceal payloads in rarely triggered paths (e.g., exception handlers). 
To ensure comprehensive coverage, \tool combines standard entry point analysis with a novel Dormant Function Analysis (DFA).

First, \tool executes standard entry points found in package metadata. For PyPI packages, this includes \texttt{setup.py}, \texttt{setup.cfg}, scripts in \texttt{pyproject.toml}, and the top-level \texttt{\_\_main\_\_.py}. For compiled binaries, \tool executes the unpacked main script and bootloaders (e.g., \texttt{pyi\_rth\_*.py}). However, high-risk payloads often reside in "dormant" functions exported as library APIs but never invoked during the installation process. To uncover these, \tool’s DFA phase operates as follows: 
During the initial run, \tool records all defined callable objects (functions and methods). 
Any callable object declared but not executed is flagged as dormant. 
Then \tool proactively invokes each dormant API in an isolated process. For class methods, \tool will invoke the initialization methods first, then invoke the corresponding dormant methods. For functions, \tool will directly invoke it.
Since these context-free invocations often lack valid arguments, our crash prevention mechanism synthesizes objects to satisfy function signatures. 
Based on our observations of the malicious packages, the injected payloads are often self-contained, so the context-free invocation is enough to expose their behaviors.

In the ultrarequest example, DFA automatically identifies and invokes the unexecuted HTTPError methods. 
\tool's instrumentation of exception blocks (e.g., SETUP\_FINALLY) ensures that even logic hidden within error handlers is explored. 
This aggressive, context-free invocation prevents sophisticated attackers from hiding payloads in rarely used components.

\section{Implementation}
\label{sec:implementation}
To demonstrate the practical utility of the execution traces produced by \tool, we implemented \toolscan, a lightweight detector that uses \tool to analyze the target program together with an LLM component (GPT-4.1-mini) to classify packages based on the resulting traces. 
In \toolscan, \tool and the LLM component serve distinct roles: \tool is the analysis engine that exposes the target application's behavior and generates structured, strace-style~\cite{straceStrace} execution traces of function calls, while the LLM component analyzes only the traces generated by \tool to determine whether a package is malicious. 
It does not guide execution, recover paths, or synthesize runtime states. 
While prior works~\cite{malguard, llmnpmmalicious} have applied LLMs to malware analysis, they are constrained by incomplete inputs from static or fragile dynamic tools. 
\toolscan instead leverages the more comprehensive and semantically rich traces produced by \tool to capture evasive, runtime-dependent behaviors. 
A key feature is the representation of the \textit{Synthetic Object}: rather than acting as empty placeholders, these objects carry the trace properties that record their origin and computational history (as detailed in Section~\ref{sec:meth:crash_prevention}). 
We explicitly prompt the LLM (Appendix~\ref{sec:app:LLMprompt}) to interpret this lineage to reconstruct data flow when concrete values are unavailable, enabling it to classify the package and provide a behavioral rationale.

The purpose of \toolscan is not to serve as the scientific contribution of this paper, but rather to provide a concrete means of evaluating the value and fidelity of the semantic insights generated by \tool's crash-free, forced-execution-based dynamic program analysis. 
Without the LLM, \tool alone would still be useful for malware analysis because it would expose the same hidden behaviors and generate execution traces for manual inspection. 
The LLM enables \toolscan to automatically and robustly identify malware at scale.

\section{Evaluation} 
\label{sec:evaluation}
We evaluate \tool and its detection version, \toolscan, focusing on three dimensions: (1) the accuracy of detection on a ground truth dataset, (2) the depth of behavioral insights enabled by our unique design, and (3) the practical utility in discovering new threats in the wild.

\subsection{Experimental Setup} 
\label{sec:eval:setup} 
All experiments were conducted on a workstation equipped with an Intel Core i9-13900 CPU and 128 GB of RAM. \tool is prototyped on top of CPython 3.7, 3.11, and 3.12. \toolscan utilizes the GPT-4.1-mini model for classification, processing the semantic execution traces generated by \tool. We utilize three distinct datasets to evaluate different aspects of our system: 
\begin{itemize}
    \item Malicious PyPI Packages (D1-Raw): A collection of 10,920 known malicious PyPI packages, sourced from established academic and security community datasets~\cite{ohm2020backstabber, pypimalregistry, ea4mp, malguard}. 

    \item Malicious Python Executables (D2): A curated set of \numDtwo real-world malicious executables collected from the wild, which were originally developed using Python. 

    \item Wild PyPI Packages (D3): A longitudinal dataset of \numtotalpypiscanned packages collected from the PyPI daily updated feed over a three-month time span. This corpus represents the real-world velocity of package publications.
\end{itemize}

\subsection{Ground Truth Dataset (D1-Raw) Calibration and De-duplication} \label{sec:eval:calibration}
Initially, we applied \toolscan to the D1-Raw dataset and manually inspected the failure cases to identify the limitations of our tool. However, the initial manual inspection of the results revealed significant flaws in this established benchmark, which has been used in prior studies~\cite{malguard, ea4mp}.

\textbf{Dataset Contamination.} We discovered substantial misclassification within the D1-Raw dataset. We identified 112 packages labeled as "malicious" that were, in fact, functionally benign. 
We found 53 packages containing only trivial \texttt{print} statements, 7 packages instructing users to install a correctly named alternative, 1 package explicitly raising an exception about incorrect installation, and 2 packages using offensive language but lacking any malicious program behavior. 
Additionally, 16 packages were empty projects without meaningful scripts, 3 packages killed processes based solely on user-provided PIDs (a potentially sensitive but not inherently malicious function), 4 used shell commands merely to display system information, and 2 created harmless "hello-world" files. 
Notably, 22 packages contained unrelated executable binary payloads within their source directories, but the Python code itself was benign and did not interact with these binaries, placing them outside the scope of Python analysis. 
Finally, 2 packages were benign tools for encoding data into media files, although they could be abused to hide malicious payloads. We suspect these misclassifications stem from automated collection methods. For instance, PyPI often purges all versions of a package if one version is malicious, leading collectors to mistakenly flag benign historical versions as malicious. We hypothesize these false positives in the ground truth stem from aggressive automated collection strategies. Furthermore, packages removed for violating community guidelines (e.g., language abuse) may have been incorrectly categorized as programmatically malicious.

\toolscan correctly identified these 112 packages as benign from a Python-based malware perspective. We reported these erroneous samples to the dataset maintainers, who have since confirmed our findings and initiated corrections.

\textbf{Semantic Redundancy.} Furthermore, we observed that D1-Raw is heavily inflated with polymorphic clones—packages that differ in file hash or variable names but possess identical control flow and logic. Evaluating on such a dataset artificially inflates performance metrics. 
To address this, we developed a simple de-duplication tool. This static analysis tool constructs a signature for each package by iterating through its Function Call Graph (FCG) and capturing the associated argument structures. Packages yielding identical FCG signatures were grouped, and only one representative sample was retained for each group. Note that this is a conservative de-duplication mechanism. The deduplication tool did not handle dynamically generated payloads, so a slight change in the encoded payload can turn semantically identical code into different code from the deduplication tool's perspective.

\textbf{The Calibrated Dataset.} After removing benign samples and de-duplicating semantically identical packages, we reduced the D1-Raw dataset to 2,961 unique, confirmed malicious samples. We refer to this as the \textbf{D1-Calibrated}. All subsequent comparative evaluations are conducted on this rigorous set to provide a fair assessment of detection capabilities.

\subsection{Comparative Detection Effectiveness}
\label{sec:eval:tool_effectivenss}
\begin{table}[h!] 
\centering 
\scriptsize
\caption{Detection Performance on Malicious Packages (D1)} 
\label{tab:results_d2} 
\begin{tabular}{l c c c} 
\toprule 
\textbf{Tool} & \textbf{Precision} & \textbf{Recall} & \textbf{F1-Score} \\ \midrule \textbf{\toolscan } & \textbf{99.1\%} & \textbf{99.9\%} & \textbf{99.5\%} \\ \midrule 

Malguard & 94.7\% & 89.5\% & 91.7\% \\ 
EA4MP & 91.3\% & 86.0\% & 88.6\% \\
GuardDog & 86.4\% & 77.0\% & 81.5\% \\
Bandit4Mal & 83.8\% & 71.8\% & 77.4\% \\ \midrule 
Package Analysis (Dynamic Only + GPT-5) & 50.0\% & 45.2\% & 47.4\% \\ 
Package Analysis (GPT-5) & 93.6\% & 83.8\% & 88.4\% \\

\bottomrule 
\end{tabular} 
\end{table}

We compared \toolscan against state-of-the-art static analysis tools including Malguard~\cite{malguard}, EA4MP~\cite{ea4mp}, GuardDog~\cite{guarddog}, Bandit4Mal~\cite{band4mal} and a hybrid(dynamic+static) analysis tool (Package Analysis~\cite{Ossfpackageanalysis}) using the Calibrated Dataset and the same amount of popular benign packages collected from PyPI following the same method used in Malguard~\cite{malguard}.

\subsubsection{Results on PyPI Packages} 
\toolscan outperformed all baselines, achieving an F1-Score of 99.5\% as shown in Table~\ref{tab:results_d2}. 
The comparative analysis highlights the limitations of existing approaches:

\textbf{Limitation of Static Analysis.} Static tools like Malguard and EA4MP struggled to detect malware that employed runtime evasions. 
Firstly, static analysis tools like GuardDog and Bandit4Mal rely on parsing the abstract syntax tree; however, the AST parsing tool is not reliable. For instance,in Bandit4mal, we observed 39 AST errors in the benign package dataset and 95 AST errors in the malicious package dataset.
Secondly, a significant portion of the dataset contained malware that concealed its payload using techniques such as base64-encoded strings passed to the eval() function. 
Instead, their rules are overfitting to such cases and causing false positives. 
Static analyzers cannot reliably identify these patterns as malicious without runtime context. 
\tool's forced execution and dynamic analysis, however, ensures that these hidden blocks are executed to expose the real payload, and its data provenance tracking reveals the flow of obfuscated data to sensitive sinks, providing the LLM with the evidence of malicious behavior. 
Moreover, static analysis tools often fail to differentiate user-controlled sensitive behaviors from attacker-controlled sensitive behaviors. 
For instance, a sensitive API call (e.g., \texttt{exec} or \texttt{subprocess.Popen}) is often flagged as malicious based on heuristics targeting obfuscation or encoding using PyArmor~\cite{githubGitHubDashingsoftpyarmor}. 
However, these patterns also exist in legitimate code protecting intellectual property. 
Conversely, if a malware author uses a novel encoding scheme not present in the static tool's rules, the detection is easily bypassed. 
\tool has better performance because it tracks the provenance of the data: it distinguishes between a user-supplied argument (benign usage) and an attacker-controlled, hardcoded payload (malicious), regardless of the obfuscation used.

\textbf{Limitations of Dynamic Analysis.} We evaluated the OSSF sandbox, which utilizes a hybrid framework aggregating both static and dynamic features. We then feed the collected features into state-of-the-art LLMs (GPT-5) for LLM-based classification. 
Our analysis suggests that while this approach generally outperforms pure static analysis, it appears to suffer from a critical fallback limitation: when the dynamic component is evaded, the system effectively degrades into a static analyzer, inheriting the fragility associated with static signatures. 
This observation is supported by our isolation of the dynamic analysis mode, which achieved an accuracy of only about 50\%, highlighting the challenges faced by traditional runtime analysis. This performance gap is driven by the prevalence of evasion techniques. 
59\% of the samples in our Calibrated Dataset employ environmental guardrails (e.g., checking for specific usernames, timestamps, or virtualization artifacts) or conceal logic behind non-standard API entry points. 
Whereas traditional dynamic analysis executes a single path and remains dormant when these checks fail, \tool's forced execution explores both branches, ensuring the malicious payload is exposed regardless of the environment, and the entry point identification forcefully invokes the naturally uncovered APIs to expose the hidden malicious behaviors.

\subsubsection{Results on Compiled Executables} 
\begin{table}[t!]
    \centering
    \footnotesize
    \caption{D2 Malware Types Categorized by \tool}
    \label{tab:malware_d2_dist}
    \begin{tabular}{lcc} 
        \toprule
        \textbf{Malware Type} & \textbf{Number} &\textbf{Percentage (\%)} \\
        \midrule
        Payload Execution & 244 &58.51 \\
        Malware Framework & 97 &23.26 \\
        Information Stealers & 28 & 6.71 \\
        Remote Access Trojans & 9 & 2.16 \\
        Adware & 9 & 2.16 \\
        CVE Exploit & 5 & 1.20 \\
        Persistence Dropper& 5 & 1.20 \\
        Cannot be executed & 5 & 1.20 \\
        Stdout/stderr Interceptor & 4 & 0.96 \\
        Keylogger & 3 & 0.72 \\
        Ransomware & 3 & 0.72 \\
        Blank & 3 & 0.72 \\
        Android Exploit & 1 & 0.24 \\
        Cloudflare Bypass & 1 & 0.24 \\
        \bottomrule
    \end{tabular}
\end{table}

D2 dataset (Python-based Executables) consists of 417 in-the-wild Python-based executables provided by an industry security partner. 
These compiled Python-based executables run independently by packaging all necessary Python bytecode files (about 319 files per executable) within themselves. 

All existing analysis tools failed entirely on this dataset. These tools are designed for Python source code and cannot interpret the packed bytecode produced by tools like PyInstaller. \tool, however, operates directly on the bytecode, allowing it to analyze the malware's logic regardless of the initial packaging, a fundamental advantage over all static methods. 
OSSF Package Analysis~\cite{Ossfpackageanalysis}, as a hybrid analysis framework, also does not support executing these binaries out of the box. 
Even if we invoke the compiled module through our manually crafted Python loader, it still fails to identify the malware constantly.
This lower performance is attributed to the prevalence of anti-analysis checks within the unpacked malware, which \tool's forced execution critically bypasses, ensuring complete behavioral exposure.

Our analysis starts from the executable's entry point and deems it malicious if \toolscan detects any malicious behavior during analysis.
We do not attempt to further differentiate individual bytecode files, since the distributed artifact encountered by the users is an executable as a whole. Table~\ref{tab:malware_d2_dist} details the diverse functionalities observed within the D2 dataset. 
These malware types are derived from the behaviors exposed in \tool traces, assisted by LLM-based labeling and then manually verified. 
The distribution was dominated by Payload Execution and Malware Frameworks, which are samples that primarily serve as framework-generated scaffolding rather than standalone payloads, followed by Information Stealers. We also identified a small subset of samples as benign and they were manually verified as broken (\textit{Cannot be executed}) or empty (\textit{Blank}).
Our analysis revealed that the Python component in these executables sometimes functions as a \textit{dropper}. 
In these cases, the Python logic contains barely malicious but sensitive functionality itself, serving only to extract and launch a secondary payload, typically a VBScript or native binary embedded in the resource section. 
\tool's execution traces still expose the underlying file extraction sequences and process creation, enabling \toolscan to correctly classify the behavior as malicious.

\subsection{Key Design Validations} \label{sec:eval:deep_analysis}
\begin{table}[t]
\centering
\footnotesize
\caption{Capabilities of Malicious Package Exposed by \tool}
\label{tab:malware_char}
\begin{tabular}{lrr}
\toprule
\textbf{Technique (Example)} & \textbf{Number} & \textbf{Percent} \\
\midrule
\multicolumn{3}{l}{\textit{\textbf{Static Analysis Evasion}}} \\
Encoding (Base64) & 1,654 & 55.84\% \\
Runtime Code Execution (exec, eval) & 2,273 & 76.76\% \\
Process Spawn (os.system, popen) & 739 & 24.96\% \\
Encryption (Fernet, AES) & 1060 & 35.79\% \\
Compression (zlib, gzip) & 167 & 5.64\% \\
Obfuscation (PyArmor, Minify) & 67 & 2.26\% \\
\midrule
\multicolumn{3}{l}{\textit{\textbf{Dynamic Analysis Evasion}}} \\
Sandbox Tool Check (QEMU) &  184 & 6.21\% \\
Debugger Tool Check (Windbg) & 206 & 6.95\% \\
Hardware Check (CPU, RAM) &  93 & 3.14\% \\
Timing Evasion (Sleep) & 345 & 13.2\% \\
Conditional Imports & 30 & 1.01\% \\
Network Check (DNS) & 218 & 7.36\% \\
OS Check (OS Type) &  486 & 16.41\% \\
Artifact Check (Telegram) & 664 & 22.42\% \\
User Check (Mouse Movement) & 46 & 1.55\% \\
\midrule
\multicolumn{3}{l}{\textit{\textbf{Persistence}}} \\
Shell Config (.bashrc) & 396 & 13.37\% \\
Windows Registry (Reg, Tasks) & 129 & 4.36\% \\
Startup folder & 97 & 3.27\% \\
Cron Job & 31 & 1.05\% \\
Anti-virus Disable (UAC) & 41	& 1.38 \% \\
Tampering & 73 & 2.46\% \\
App. Injection (Browser Ext.) & 47 & 1.59\% \\
\midrule
\multicolumn{3}{l}{\textit{\textbf{C2 Domain}}} \\
Custom Domain & 88 & 2.97\% \\
Web/Cloud Hosting & 122 & 4.12\% \\
Messaging (Discord) & 201 & 6.79\% \\
Repo. (GitHub) & 126 & 4.25\%\\
OAST (Burp Collab.) & 55 & 1.86\% \\
File Storage (Dropbox) & 43 & 1.45\% \\
Text Storage (Pastebin) & 106 & 3.58\% \\
\midrule
\multicolumn{3}{l}{\textit{\textbf{Malicious Code Entry}}} \\
Auto-Trigger (Import/Install) & 2,408 & 82.32\% \\
User-Trigger (Specific API Call) & 564 & 19.04\% \\
\bottomrule
\multicolumn{3}{r}{\footnotesize *One package may use multiple techniques.}
\end{tabular}
\end{table}

To demonstrate that our high detection rates result from our design choices, we conducted a detailed analysis of execution traces generated by \tool from D1-calibrated data, as shown in Table~\ref{tab:malware_char}.

\textbf{Evasion and Forced Execution.} We categorized the evasion techniques employed by the samples to validate our design choices. We found that over 1,654 packages (55.8\%) utilize techniques such as encoding to conceal malicious logic from static analysis tools. Consequently, static analysis is often limited to identifying suspicious code smells rather than faithfully reporting confirmed malicious behaviors. Furthermore, more than 22.4\% of packages utilize conditional checks to shield their payloads from detection by standard sandboxes or debuggers. \tool's \textit{Forced Execution} proved to be a strictly necessary component, enabling the system to dynamically expand runtime-generated payloads and expose malicious behaviors hidden behind these conditional guards.

\textbf{Process Merging.} To quantify the impact of our post-dominator path merging strategy, we simulated a naive forced execution without branch pruning on the D1-Calibrated dataset. Without merging, the analysis rapidly succumbs to the path explosion problem, exhausting available memory. 
Even under a conservative estimate, with all additional processes terminated when the function returns and excluding outliers that spawned over 100,000 processes, the naive approach generated an average of 955 processes per package. In contrast, \tool's merging mechanism spawned a lean average of just 44 processes. 
The necessity of merging is most critical for the 73 packages that exceeded the 100,000-process threshold in the naive simulation. 
Our analysis revealed that these are often malware built on top of complex, benign packages, where attackers hide malicious logic deep within legitimate, branching-heavy code. 
This structure poses an insurmountable challenge for existing forced-execution tools. 
The most extreme example we encountered was \texttt{ressy}, a package masquerading as a calculator which contained 10,405 \texttt{if} statements within a single function. 
While this structure causes an unmanageable number of states for existing forced execution based techniques, \tool's process merging mechanism successfully consolidates these divergent paths at their post-dominator, ensuring the analysis remains computationally tractable by having at most 3 processes running in parallel.

\textbf{Resilient Crash Prevention.} A major challenge in analyzing the malware for existing forced-execution based techniques is that they will terminate prematurely before actually hitting the malicious code. 
The most common crash in our dataset is caused by the dead server problem. 
To evaluate \tool's \textit{Resilient Crash Recovery}, we extracted all Command and Control (C2) domains from our traces and tested their liveness. Only 29 of the C2 servers were still accessible, 
while the vast majority were offline or seized. 
In a traditional sandbox or a fragile forced-execution engine (like MalMax~\cite{malmax} or FV8~\cite{FV8}), the network socket error resulting from a dead C2 would terminate the execution thread immediately. 
However, our traces show that \tool successfully synthesized objects, allowing the execution to proceed. 
Crucially, this allowed us to observe subsequent behaviors, such as persistence mechanisms (e.g., modifying \texttt{.bashrc}), which only occur after the C2 communication logic has been implemented. 
This demonstrates that our crash recovery mechanism is not merely a stability feature, but a fundamental requirement for comprehensive behavioral modeling.

\textbf{Entry Point Identification.} 
A critical limitation in detecting evasive malware is the inability of dynamic analyzers to identify and exercise all viable entry points. As shown in Section~\ref{sec:eval:tool_effectivenss}, conventional tools typically rely on shallow exploration, executing only code triggered during installation or import. 
However, attackers are increasingly concealing malicious logic within library APIs, keeping payloads dormant until they are invoked by a dependent application. 
Our dataset characterization identifies two primary triggering strategies for the malicious code specifically: \textit{Auto-Triggers} (81.3\% of samples), which exploit standard hooks like \textit{setup.py} for immediate execution, and \textit{User-Triggers} (19.04\%), where malicious code embedded in callable objects remains inert until explicitly invoked. 
The discovery of user-triggered samples highlights the blind spot in existing scanners. 
\tool addresses this by identifying and forcefully executing these dormant objects, utilizing its semantic-preserving mode to generate the rich execution traces required for detection.

\textbf{Coverage Analysis} To quantify \tool's impact on code visibility, we measured the additional coverage achieved by executing statements and functions directly within the instrumented interpreter. Specifically, we record the executed bytecode for each interpreter frame during analysis and map it to statements or functions using metadata stored in the associated \texttt{PyCodeObject}. 
To measure the contribution of each mechanism, we compare executions of the same package while selectively enabling or disabling individual \tool features, including forced-execution, resilient crash recovery, and dormant function analysis. 

Under this setup, DFA naturally emerged as the most significant driver of coverage, expanding the total volume of executed statements by 47.01\%. By extending coverage beyond standard entry points to explicitly invoke all defined callable objects, DFA exposes dormant malicious APIs that are present in a package but would otherwise remain invisible to ordinary execution.
This is particularly helpful, as Python malware developers frequently inject malicious logic into the APIs of existing benign packages.
Beyond DFA, \tool's features further expanded visibility.
For branches protected by evasion conditions where the guarded code is syntactically present but unreachable in a conventional single-path run, Forced-Execution makes both sides visible in the trace, exposing an average of 20.45 additional statements and 17.41 additional function calls on packages using evasion techniques. 
Similarly, for paths that would normally terminate due to fatal crashes caused by missing dependencies or dead C2 servers, Crash Recovery allows the analysis to continue. 
This proved critical for packages encountering runtime failures, unlocking an average of 93.47 additional statements and 70.53 additional function calls per package.
Additionally, \tool successfully captures behaviors that typically bypass static inspection. For runtime-generated payloads, where static analysis often sees only encoded strings or \texttt{exec}/\texttt{eval} payloads, \tool exposes the materialized downstream behavior at runtime. 
Likewise, for compiled Python executables where source-level inspection is unavailable, \tool analyzes the bytecodes and directly exposes the malicious behaviors.
Ultimately, the additional coverage reported here corresponds to semantically meaningful behaviors exposed during execution, rather than merely a larger amount of source text being parsed. This is reinforced by the fact that the overall percentage gain in function calls (8.9\%) exceeds that of statements (5.0\%), confirming that the underlying logic exposed by \tool is highly action-dense and rich in function invocations.

\subsection{Real-World Threat Discovery}
To evaluate the real-world utility of \toolscan, we deployed it to scan a live stream of packages uploaded to PyPI over 90 days beginning from August 1, 2025.
\toolscan reports \numalert packages as malicious in total. We first manually verify the report's correctness and confirmed \numnewmalware previously unknown malicious packages.
These packages had cumulatively been downloaded \numnewmalwaredownload times by the time of detection, highlighting the severity of the threat. 

We further investigate the root cause of the 10 false-positive cases and find that they primarily stemmed from interpretive inconsistencies in the LLM classifier rather than inaccuracies in the underlying execution trace generated by \tool.
Eight cases are caused by applications performing aggressive actions, such as benign IT automation scripts used to terminate processes or modify network configurations. 
While the LLM typically utilized \tool's provenance data to correctly identify these as user-controlled (and thus benign), it occasionally misclassified them as malware despite the clear evidence of user intent. In these eight cases, the LLM correctly analyzed the traces based on the reply's reason field, but mistakenly output `malicious' in the LLM's reply's result field due to the possible hallucination.
Others are caused by the sample usage code of penetration testing tools invoked by \tool's aggressive \textit{Dormant Function Analysis}. By proactively invoking all callable objects, the system executed dangerous sample code with hard-coded targets embedded in penetration-testing libraries. 
Although the repository documentation contextualized these snippets as educational, \tool faithfully executed and reported their inherently risky logic.
Unlike false positives in static analysis, which often result from benign obfuscation, these alerts are triggered by \textit{true} risky behaviors faithfully reported by \tool. 
Consequently, they still provide high-value signals to security auditors, even when the underlying intent is benign.

We also investigated discrepancies in which packages were quarantined and removed by the PyPI security team but were considered benign by \toolscan. 
This analysis revealed a specific class of false negatives involving brute-forcing and enumeration tools. 
For example, we identified packages that iterate over internal Instagram or AWS APIs to validate account existence. 
While \tool correctly exposed these high-volume network behaviors in the execution trace, the LLM classifier did not categorize the enumeration as malicious from the program's perspective, likely interpreting it as grayware or testing activity.

Beyond active malware, our semantic traces also find other issues. 
We also have 7 packages containing hard-coded secrets embedded directly in the source code.
We reported these to the PyPI security team, and they notified the respective developers. 
This demonstrates \tool's utility could potentially extend beyond malware detection to broader supply chain risk assessment. False positives encountered and the case study discussion are available in Appendix~\ref {sec:app:casestudy}.

\section{Discussions}
\label{sec:discuss}

\textbf{Evading \tool} Conditional-dependent malware may use encryption to obfuscate the payload to evade detection. The encryption key may not be available at the time of analysis (e.g., request a decryption key from an abandoned server). 
\tool is not able to expose the malicious behaviors in the encrypted payloads if the key is unavailable. However, such practice is rare, and we still identified several such instances in the known malicious packages as the data flow of retrieving data remotely to decrypt a payload feeding into dynamic code execution APIs is suspicious based on \toolscan.

\noindent
\textbf{Handling Infinite Loop.}
To prevent analysis stagnation caused by infinite loops or excessive recursion, \tool enforces an empirically derived execution threshold 200 in the prototype. 
This mechanism limits the number of iterations, forcing the analysis to break the cycle and proceed to the subsequent control flow block once the limit is reached. 
This design remains fully configurable to accommodate varying performance constraints and aligns with the loop-bounding strategies established in prior forced-execution frameworks such as FV8~\cite{FV8} and MalMax~\cite{malmax}.

\noindent
\textbf{Performance of \tool} While \tool is primarily architected for offline security vetting rather than runtime monitoring, operational efficiency is essential for large-scale supply chain protection. 
On average, \tool completes analysis in roughly 75 seconds per package. This throughput ensures threats are detected before widespread adoption.
Performance bottlenecks are typically observed in large libraries where the \textit{Dormant Function Analysis} module attempts exhaustive API invocation. 
To mitigate this, future optimizations could leverage initial Function Call Graph (FCG) analysis to prioritize high-coverage API entry points, thereby reducing the computational redundancy caused by blind invocation.

\noindent
\textbf{Scope of \tool} \tool targets the Python execution layer by instrumenting the Python Virtual Machine, covering both source code and compiled bytecode. 
Consequently, the internal control flow of native extensions (e.g., C/C++ modules) executing directly on the host CPU and data-driven attacks interpreted by format-specific virtual machines remain outside our direct instrumentation scope. 
However, \tool maintains visibility into boundary interactions and Python-level behaviors triggered, capturing API calls and arguments passed to these modules. 
For coverage on C/C++ modules, \tool can function within a hybrid pipeline, resolving managed Python logic while offloading native components to binary forced-execution engines such as X-Force~\cite{xforce}.

\section{Related Work}
\label{sec:related_work}

Our work focuses on malware analysis, advancing beyond established static and dynamic approaches to address their inherent limitations in the face of modern evasive threats.

\textbf{Static Analysis.}
Static analysis remains popular for its scalability. Early approaches like Bandit4Mal~\cite{band4mal}, OSSGadget~\cite{OSSGadget}, HERCULE~\cite{HERCULE}, and GuardDog~\cite{guarddog} rely on AST parsing and pattern matching (e.g., YARA~\cite{yara}, Semgrep~\cite{semgrep}) to flag suspicious APIs. 
To reduce false positives, recent works leverage machine learning and graph analysis. CEREBRO~\cite{CEREBRO} and EA4MP~\cite{ea4mp} utilize transformer models (BERT~\cite{devlin2019bert}) to interpret API sequences, while MalGuard~\cite{malguard} and MalWuKong~\cite{malwukong} employ graph-theoretic approaches to filter irrelevant calls and track data flows. Others, such as PyRadar~\cite{pyradar}, detect discrepancies between PyPI artifacts and their corresponding source repositories. HERCULE~\cite{HERCULE} leverages a multi-staged static analysis beginning from artifact integrity to transitive dependency analysis to flag malicious packages.
While scalable, these methods are fundamentally limited by their lack of runtime context. They are easily blinded by dynamic code loading and cannot analyze compiled Python binaries or packed executables. 
Unlike these tools, \tool operates at the bytecode interpreter, enabling it to bypass obfuscation and analyze both source and compiled threats with equal fidelity.

\textbf{Dynamic Analysis.}
Dynamic analysis observes behavior in controlled environments to capture runtime payloads. Frameworks like MalOSS~\cite{maloss} combine static and dynamic features, while OSCAR~\cite{oscar} introduces fuzzing to trigger functions not directly executed in the installation or import stage.
Conventional dynamic tools suffer from fragility. They typically monitor a single execution path and terminate upon encountering environmental guardrails or fatal errors. 
In contrast, \tool employs \textit{Resilient Crash Recovery} to synthesize objects, allowing analysis to proceed past fatal errors, and uses \textit{Dormant Function Analysis} to systematically invoke unexecuted APIs without requiring valid inputs.

\textbf{Forced Execution.}
Forced execution was pioneered to expose logic hidden behind conditional triggers. 
X-Force~\cite{xforce} demonstrated this for binaries, while J-Force~\cite{j-force} and FV8~\cite{FV8} applied it to JavaScript engines to detect evasive browser extensions. 
Similarly, MalMax~\cite{malmax} introduced multi-aspect execution for PHP.
Prior interpreted-language approaches have significant drawbacks for Python malware. FV8 relies on fragile AST instrumentation, making it incompatible with compiled bytecode, and MalMax lacks robust crash recovery, halting analysis when network dependencies fail. 
\tool is the first tool designed for Pythonic malware, enabling robust bytecode-level forcing and scope-aware state sharing to handle complex, state-dependent malware that defeats prior art.

\textbf{LLM-Based Detection.}
Recent works like SpiderScan~\cite{SpiderScan} and SocketAI~\cite{llmnpmmalicious} leverage Large Language Models (LLMs) to classify behaviors. 
RuleLLM~\cite{rulellm} uses LLMs to refine detection rules.
PyGuard~\cite{PYGUARD} uses LLMs to abstract raw code into semantic behavioral sequences to reduce false positives.
The effectiveness of LLMs is constrained by the quality of their input context. 
By providing complete, crash-free execution traces replete with data provenance, \tool enables LLM-based detectors (like \toolscan) to reason about behavior with significantly higher accuracy than systems relying on static features or partial logs.

\textbf{Interpreter-based symbolic and concolic execution.}  Prior work has explored interpreter-based symbolic/concolic execution for dynamic web applications and interpreted languages~\cite{symphp,Chef,wassermann2008dynamic}. These systems share our view that the interpreter is the right layer for faithfully handling dynamic language features. 
However, they primarily focus on feasible execution and test input generation via symbolic reasoning, whereas \tool focuses on revealing hidden behaviors even in hostile or incomplete analysis environments.

\section{Conclusion}
\label{sec:conclusion}
The rapid expansion of the Python ecosystem has been shadowed by a parallel rise in sophisticated, evasive threats that challenge traditional analysis paradigms. 
As adversaries increasingly employ runtime obfuscation, environmental guardrails, and compilation to conceal malicious logic, existing static and dynamic analysis tools have struggled to maintain visibility. 
In this work, we presented \tool, a robust forced-execution engine integrated directly into the CPython interpreter designed to highlight these blind spots. 
Unlike prior approaches that are often fragile or limited to source code, \tool provides a unified and robust platform capable of analyzing both PyPI source packages and compiled binaries. 
Through novel mechanisms including resilient crash recovery, semantic-preserving path merging, and dormant function analysis \tool systematically exposes malicious behaviors that remain hidden to conventional analysis.

The efficacy of this methodology was validated through the development of \toolscan, a proof-of-concept detector leveraging the rich execution traces. 
Our evaluation demonstrated solid detection performance against state-of-the-art baselines and identified critical flaws in established ground-truth datasets, resulting in the correction of 112 misclassified benign packages. 
Most significantly, a live deployment on the Python Package Index uncovered \numnewmalware previously unknown malicious packages that had already accrued over \numnewmalwaredownload downloads. 
These findings underscore the urgent need to adopt resilient, exhaustive dynamic analysis techniques to secure the software supply chain against the next generation of Python-based threats.

\section{Ethical Considerations}
\label{sec:ethics}
Our research involves the analysis of live malicious software and the large-scale scanning of a public software repository. To uphold ethical standards and minimize potential harm to the software ecosystem and third parties, we adhered to the principles of responsible security research throughout our study.

We followed a responsible disclosure process for all findings:
\begin{itemize}
    \item \textbf{Malicious Packages:} Upon identifying previously unknown malicious packages on PyPI, we immediately reported them to the PyPI security team using the official reporting channels. We provided evidence from our execution traces to facilitate their rapid removal and quarantine. The PyPI security team has confirmed all of our findings and removed the reported packages.
    \item \textbf{Dataset Calibration:} We disclosed the misclassified benign packages found in the ground-truth dataset to the dataset maintainers to improve the quality of future research benchmarks.
    \item \textbf{Leaked Secrets:} For packages containing hardcoded secrets (e.g., AWS keys, Slack tokens) that were not malicious but violated security hygiene, we notified the PyPI security team. 
\end{itemize}

Our large-scale scanning of the PyPI repository was designed to be non-intrusive.
We utilized the public RSS feed to trigger analysis and respected the repository's API rate limits to ensure our scanning imposed no significant load on PyPI's infrastructure or disrupted service for legitimate users.

\printbibliography
\appendix
\section{Appendix}
\label{sec:appendix}

\subsection{Case Study}
\label{sec:app:casestudy}
\subsubsection{Identified New Malware Distribution Campaign}
This case study examines a PyPI supply-chain-attack campaign  we name \texttt{PyGet}, which masqueraded as a command-line utility for downloading and executing remote binaries. The packages distributed in Campaign have more than 3,000 times downloads in one-week campaign period.
This campaign highlights two critical challenges in supply chain security: the difficulty of distinguishing malicious behavior from benign dual-use functionality and the prevalence of trivial, high-volume attacks that evade hash-based blocklisting. 
We demonstrate how \tool's forced execution analysis' trace overcomes the limitations of static API call graph analyzers like Malguard~\cite{malguard} in this scenario.

\paragraph{The Attack: A Deceptive Utility}
The packages used in the campaign advertised themselves as a simple tool: a user could provide a URL via a command-line argument, and the script would download the binary from that URL and execute it. 
However, embedded within obfuscated code, the package contained a secondary, covert function. 
In addition to the user-requested operation, it would decode a hardcoded, base64-encoded URL, download a malicious binary from this attacker-controlled endpoint, and execute it in the background. 
The attackers employed simple but effective obfuscation, using tools like PyArmor~\cite{githubGitHubDashingsoftpyarmor} to hide the API calls for the malicious download (\texttt{requests.get}) and execution (\texttt{subprocess.run}).

\paragraph{Limitation of Static Analysis}
We first analyzed \texttt{PyGet} using a methodology similar to Malguard. 
The obfuscation posed a significant challenge to its static analyzer, preventing the extraction of a complete and accurate API call graph, thus leading to a false negative. 
If the static-based tool blindly treat all packages packed with PyArmor as malware, it will cause large volumes of false positives as we have encounter benign packages using PyArmor as well.

To further investigate, we manually de-obfuscated the code and re-analyzed it. The resulting API call graph correctly identified a \texttt{requests.get} $\rightarrow$ \texttt{subprocess.run} pattern, which is inherently suspicious. 
However, this is where a more fundamental weakness emerged. We compared this graph to that of a popular, benign open-source utility that performs the exact same function (downloading and running a user-specified binary). 
Their API call graphs were identical. 
Because Malguard's graph representation lacks data-flow information. 
Specifically, the provenance of function arguments, it cannot distinguish between them. 
It does not know that one URL is supplied by the user at runtime while the other is hardcoded by an attacker. 
Consequently, this approach would flag the benign package as malicious, resulting in a critical false positive.

\paragraph{Disambiguation via Object Trace Property}
In contrast, \tool's forced execution engine successfully navigated the obfuscation and produced a high-fidelity execution trace that captured the package's true behavior. The trace clearly revealed two distinct execution flows:
\begin{enumerate}
    \item \textbf{Benign Flow:} A data path originating from a user-controlled source (\texttt{sys.argv}) was passed to \texttt{requests.get} and subsequently executed.
    \item \textbf{Malicious Flow:} A hardcoded, base64-encoded string was decoded into a URL, passed to a second \texttt{requests.get} call, and the resulting file was executed via \texttt{subprocess.run}.
\end{enumerate}
This trace provided the crucial semantic context that the static call graph lacked. The presence of the second, non-user-initiated flow with a hardcoded URL is a powerful heuristic for malice. \tool correctly flagged \texttt{PyGet} as malicious while correctly identifying the benign utility as safe, as its trace only contained the first flow.

\paragraph{Campaign Identification}
The attacker's campaign did not stop with a single upload. 
Over the one week, we detected 11 different packages uploaded to PyPI under various names (e.g., \texttt{url-retriever}, \texttt{py-fetch-tool}, \texttt{quickget}). 
These packages contained the same malicious logic, altered only by changing variable names or comments. 
While each of the 11 instances had a unique file hash, \tool's trace-based similarity analysis immediately identified that they all produced semantically identical execution traces. 
This allowed us to group these disparate alerts into a single, coordinated campaign, demonstrating \tool's effectiveness in identifying polymorphic, low-effort attacks that would otherwise appear as unrelated threats to traditional, signature-based systems.

\subsubsection{Artificially Inflated Malware}
Although large language models (LLMs) have proven increasingly effective in identifying malicious code patterns, attackers are adapting, developing evasive techniques specifically designed to exploit the limitations of these models. 
We identified a malicious package on PyPI that pretends to be a calculator. 
It highlights the need for robust analysis primitives like those provided by \tool.
\begin{listing}[!ht]
\begin{minted}
[
frame=lines,
framesep=2mm,
baselinestretch=1.2,
fontsize=\scriptsize,
linenos,
numbersep=1pt %
]
{python}
if num1 == 1 and num2 == 1 and operation == '+':
    return 2
# ...More hardcoded results
elif num1 == 100:
    do_malicious()
#... More hardcoded results
elif num1 == 9999 and num2 == 9999 and operation == '+':
    return 19998
\end{minted}
\caption{Malware Sample: Calculator}
\label{listing:antillm}
\end{listing}

As shown in Listing~\ref{listing:antillm}, the package masquerades as a high-performance calculator. 
Its artificially inflated size brings a benefit of exceeding the context window limitations inherent in LLMs. 
Instead of performing actual arithmetic, the malware contains thousands of lines of hardcoded if/elif/else statements that map specific input pairs directly to pre-computed results. 
Buried deep within this expansive conditional structure is a single branch, activated by a pre-defined input (e.g., first input is 100), which triggers an obfuscated malicious payload.
This design presents multiple challenges: Anti-LLM Evasion: The excessive length of the repetitive conditional blocks floods the LLM's context window. 
Whether analyzing source code or even a partial execution trace, the LLM loses the ability to connect the specific trigger input to the distant payload. 
Path Explosion: The sheer number of branches would overwhelm less sophisticated forced-execution engines such as FV8~\cite{FV8} and MalMax~\cite{malmax}, making exhaustive exploration computationally infeasible. 
Naive Dynamic Analysis is unlikely to hit the precise input entry point required to activate the malicious path.

\tool's integrated design systematically overcomes these obstacles. 
Dormant Function Analysis identifies \texttt{calculate()} as a target and forcefully invokes it, assigning \textit{synthetic objects} if needed. 
Forced execution guarantees exploration of all branches, including the malicious ones. 
Crucially, post-dominator path merging prevents path explosion by consolidating the outcomes of the thousands of benign branches, making the analysis tractable. 
The result is a complete, yet manageable, execution trace that captures the critical sequence: invocation of calculate, forced entry into the malicious branch, and the subsequent sensitive calls (os.getenv, requests.post) in the obfuscated payload. 
While the raw code obscures the threat from an LLM, the focused, semantically rich trace generated by \tool provides \toolscan's LLM backend with the precise context needed to accurately identify the malicious behavior. This case demonstrates \toolscan's resilience against evasions targeting both traditional dynamic analysis and modern LLM-based techniques.

\subsection{Open Science}
We follow the principles of open science and reproducibility. 
To facilitate future research and verification of our results, we have released our complete artifact, which includes the \tool source code, the data preprocessing scripts, the datasets used in our evaluation, and our reproduction of related works.
The artifact is available at: \url{https://github.com/EmsecCispa/PyFEX}.

\subsection{Generative AI Usage}
In the preparation of this work, we utilized Large Language Models (specifically GPT-4.1-mini and GPT-5) to assist with grammar checking and improving clarity. 
These models also served as integral components of our experimental methodology, classifying execution traces within the \toolscan artifact and OSSF Package Analysis. 
Additionally, GPT-5-codex was employed during the early stages of prototype development to design the project's code architectural skeleton, write test cases and implement dataset pre-processing scripts. 
All AI-generated text and code were manually verified by the authors for accuracy, completeness, and adherence to the underlying scientific data.

\begin{listing*}[h!]
    
\begin{minted}[frame=lines, framesep=2mm, baselinestretch=1.2, fontsize=\scriptsize, breaklines=true]{markdown}
You are an expert malware analyst specializing in Python software supply chain security. 
Your task is to analyze the provided execution trace of a Python package to determine 
if it exhibits malicious behavior.
# Input Description
The input is a structured log of function calls, methods, and arguments captured 
during the dynamic execution of the package. 
# Critical Instruction on Synthetic Objects
The trace may contain Synthetic Objects generated by the analysis engine to recover from runtime crashes (e.g., offline servers, missing dependencies). You must treat the string representation of these objects not as literal strings, but as symbolic provenance properties.
Format: These properties describe the object creation and consumption history (e.g., `ADD(Dummy_A, 1)`).
Task: You must trace the lineage of these objects to understand the semantic context of the data flow. 
#  Output Format
Provide your analysis in the following JSON format:
{
  "result": "Malicious" | "Benign",
  "confidence": "High" | "Medium" | "Low",
  "reason": "A concise explanation citing specific behaviors and data flows from the trace that justifies the verdict."
}
# Execution Trace
[INSERT EXECUTION TRACE HERE]
\end{minted}
\caption{LLM Prompt}
\label{listing:llm_prompt}
\end{listing*}

\begin{table*}[h!]
\caption{Bytecodes Instrumented by \tool}
\label{tab:pyfex-bytecodes-310}
\footnotesize
\begin{tabular}{@{}l p{0.82\linewidth}@{}}
\toprule
\textbf{Type} & \textbf{Bytecodes} \\
\midrule
Forced Execution
& \texttt{POP\_JUMP\_IF\_FALSE}, \texttt{POP\_JUMP\_IF\_TRUE},
\texttt{JUMP\_IF\_FALSE\_OR\_POP}, \texttt{JUMP\_IF\_TRUE\_OR\_POP},
\texttt{JUMP\_IF\_NOT\_EXC\_MATCH}, \texttt{JUMP\_ABSOLUTE},
\texttt{FOR\_ITER},
\texttt{SETUP\_FINALLY}, \texttt{SETUP\_WITH}, \texttt{SETUP\_ASYNC\_WITH},
\texttt{WITH\_EXCEPT\_START}, \texttt{BEFORE\_ASYNC\_WITH},
\texttt{RAISE\_VARARGS},
\texttt{YIELD\_VALUE},
\texttt{GET\_AITER}, \texttt{GET\_ANEXT}, \texttt{GET\_AWAITABLE},
\texttt{MATCH\_MAPPING}, \texttt{MATCH\_SEQUENCE},
\texttt{MATCH\_KEYS}, \texttt{MATCH\_CLASS}. \\
\addlinespace
Crash Recovery
& \texttt{UNARY\_POSITIVE}, \texttt{UNARY\_NEGATIVE},
\texttt{UNARY\_NOT}, \texttt{UNARY\_INVERT},
\texttt{BINARY\_POWER}, \texttt{BINARY\_MULTIPLY},
\texttt{BINARY\_MATRIX\_MULTIPLY}, \texttt{BINARY\_TRUE\_DIVIDE},
\texttt{BINARY\_FLOOR\_DIVIDE}, \texttt{BINARY\_MODULO},
\texttt{BINARY\_ADD}, \texttt{BINARY\_SUBTRACT},
\texttt{BINARY\_SUBSCR}, \texttt{BINARY\_LSHIFT}, \texttt{BINARY\_RSHIFT},
\texttt{BINARY\_AND}, \texttt{BINARY\_XOR}, \texttt{BINARY\_OR},
\texttt{BUILD\_TUPLE}, \texttt{BUILD\_LIST}, \texttt{BUILD\_STRING}, \texttt{BUILD\_SLICE}, \texttt{LIST\_TO\_TUPLE}, \texttt{LIST\_APPEND},
\texttt{INPLACE\_POWER}, \texttt{INPLACE\_MULTIPLY},
\texttt{INPLACE\_MATRIX\_MULTIPLY}, \texttt{INPLACE\_TRUE\_DIVIDE},
\texttt{INPLACE\_FLOOR\_DIVIDE}, \texttt{INPLACE\_MODULO},
\texttt{INPLACE\_ADD}, \texttt{INPLACE\_SUBTRACT},
\texttt{INPLACE\_LSHIFT}, \texttt{INPLACE\_RSHIFT},
\texttt{INPLACE\_AND}, \texttt{INPLACE\_XOR}, \texttt{INPLACE\_OR},
\texttt{COMPARE\_OP}, \texttt{CONTAINS\_OP},
\texttt{LOAD\_FAST}, \texttt{LOAD\_NAME}, \texttt{LOAD\_GLOBAL},
\texttt{LOAD\_DEREF}, \texttt{LOAD\_CLASSDEREF},
\texttt{LOAD\_ATTR}, \texttt{LOAD\_METHOD},
\texttt{STORE\_NAME}, \texttt{STORE\_GLOBAL}, \texttt{STORE\_ATTR},
\texttt{STORE\_SUBSCR},
\texttt{DELETE\_NAME}, \texttt{DELETE\_GLOBAL}, \texttt{DELETE\_FAST},
\texttt{DELETE\_DEREF}, \texttt{DELETE\_ATTR}, \texttt{DELETE\_SUBSCR},
\texttt{UNPACK\_SEQUENCE}, \texttt{UNPACK\_EX},
\texttt{GET\_ITER}, \texttt{FOR\_ITER}, \texttt{GET\_AWAITABLE},
\texttt{CALL\_FUNCTION}, \texttt{CALL\_FUNCTION\_KW},
\texttt{CALL\_FUNCTION\_EX}, \texttt{CALL\_METHOD},
\texttt{BUILD\_SET}, \texttt{BUILD\_MAP}, \texttt{BUILD\_CONST\_KEY\_MAP},
\texttt{LIST\_EXTEND}, \texttt{SET\_UPDATE},
\texttt{DICT\_MERGE}, \texttt{DICT\_UPDATE},
\texttt{FORMAT\_VALUE},
\texttt{IMPORT\_NAME}, \texttt{IMPORT\_FROM}, \texttt{IMPORT\_STAR},
\texttt{SETUP\_WITH}, \texttt{WITH\_EXCEPT\_START},
\texttt{GET\_LEN},
\texttt{MATCH\_MAPPING}, \texttt{MATCH\_SEQUENCE},
\texttt{MATCH\_KEYS}, \texttt{MATCH\_CLASS},
\texttt{COPY\_DICT\_WITHOUT\_KEYS}. \\
\bottomrule
\end{tabular}
\end{table*}

\subsection{LLM Prompt}
\label{sec:app:LLMprompt}
The prompt we used to query the LLM component of \toolscan is in Listing~\ref{listing:llm_prompt}.

\subsection{Instrumented Bytecodes}
Table~\ref{tab:pyfex-bytecodes-310} shows the bytecodes instrumented by \tool's prototype.

\FloatBarrier

\end{document}